\documentclass[amsmath,amssymb,aps,prd,superscriptaddress,nofootinbib,preprint,preprintnumbers,longbibliography,a4paper,bm]{revtex4-2}
\usepackage{geometry}
\usepackage[mathscr]{euscript}
\usepackage{mathtools}
\usepackage[utf8]{inputenc}
\usepackage[T1]{fontenc}
\usepackage{verbatim}
\usepackage[normalem]{ulem}
\usepackage{graphicx}
\usepackage{color}
\usepackage{xcolor}
\usepackage{physics}
\usepackage{booktabs}
\usepackage{diagbox}
\usepackage{tabularx}
\usepackage{appendix}
\usepackage{slashed}
\usepackage{tikz}
\usetikzlibrary{angles,quotes}
\usetikzlibrary{decorations.pathmorphing,decorations.markings,arrows.meta}
\usepackage[colorlinks=true,citecolor=blue,linkcolor=purple,urlcolor=purple]{hyperref}
\usepackage[english]{babel}
\usepackage[nottoc]{tocbibind}
\usepackage{bbold}
\usepackage{adjustbox}
\usepackage{multirow}
\usepackage{subfigure}
\newcommand{\Op}{\mathcal{O}}
\newcommand{\WC}{\mathcal{C}}
\newcommand{\lt}{\lambda_t}

\newcommand{\oplqone}{\Op_{\ell q}^{(1)}}
\newcommand{\oplqthree}{\Op_{\ell q}^{(3)}}
\newcommand{\opld}{\Op_{\ell d}}
\newcommand{\opledq}{\Op_{\ell edq}}
\newcommand{\oped}{\Op_{ed}}
\newcommand{\opqe}{\Op_{qe}}
\newcommand{\opphid}{\Op_{\phi d}}
\newcommand{\opphiqone}{\Op_{\varphi q}^{(1)}}
\newcommand{\opphiqthree}{\Op_{\varphi q}^{(3)}}
\newcommand{\opdw}{\Op_{dW}}
\newcommand{\opdb}{\Op_{dB}}

\newcommand{\wclqone}{\WC_{\ell q}^{(1)}}
\newcommand{\wclqthree}{\WC_{\ell q}^{(3)}}
\newcommand{\wcld}{\WC_{\ell d}}
\newcommand{\wcledq}{\WC_{\ell edq}}

\newcommand{\wcqe}{\WC_{qe}}
\newcommand{\wcphid}{\WC_{\phi d}}
\newcommand{\wcphiqone}{\WC_{\varphi q}^{(1)}}
\newcommand{\wcphiqthree}{\WC_{\varphi q}^{(3)}}
\newcommand{\wcdw}{\WC_{dW}}
\newcommand{\wcdb}{\WC_{dB}}
\newcommand{\wcdgamma}{\WC_{d\gamma}}

\newcommand{\wclqoneMOO}{[\WC_{\ell q}^{(1)}]_{2211}}
\newcommand{\wclqoneMOT}{[\WC_{\ell q}^{(1)}]_{2212}}
\newcommand{\wclqoneMTT}{[\WC_{\ell q}^{(1)}]_{2222}}
\newcommand{\wclqthreeMOO}{[\WC_{\ell q}^{(3)}]_{2211}}

\newcommand{\wclqthreeMTT}{[\WC_{\ell q}^{(3)}]_{2222}}
\newcommand{\wcldMOT}{[\WC_{\ell d}]_{2212}}
\newcommand{\wcledqMOO}{[\WC_{\ell edq}]_{2211}}
\newcommand{\wcledqMOT}{[\WC_{\ell edq}]_{2212}}
\newcommand{\wcledqMTO}{[\WC_{\ell edq}]_{2221}}
\newcommand{\wcedMOT}{[\WC_{ed}]_{2212}}
\newcommand{\wcqeMOO}{[\WC_{qe}]_{1122}}
\newcommand{\wcqeMOT}{[\WC_{qe}]_{1222}}
\newcommand{\wcqeMTT}{[\WC_{qe}]_{2222}}
\newcommand{\wclqoneEOO}{[\WC_{\ell q}^{(1)}]_{1111}}
\newcommand{\wclqoneEOT}{[\WC_{\ell q}^{(1)}]_{1112}}
\newcommand{\wclqoneETT}{[\WC_{\ell q}^{(1)}]_{1122}}
\newcommand{\wclqthreeEOO}{[\WC_{\ell q}^{(3)}]_{1111}}
\newcommand{\wclqthreeEOT}{[\WC_{\ell q}^{(3)}]_{1112}}
\newcommand{\wclqthreeETT}{[\WC_{\ell q}^{(3)}]_{1122}}
\newcommand{\wcldEOT}{[\WC_{\ell d}]_{1112}}
\newcommand{\wcledqEOO}{[\WC_{\ell edq}]_{1111}}
\newcommand{\wcledqEOT}{[\WC_{\ell edq}]_{1112}}
\newcommand{\wcledqETO}{[\WC_{\ell edq}]_{1121}}
\newcommand{\wcedEOT}{[\WC_{ed}]_{1112}}
\newcommand{\wcqeEOO}{[\WC_{qe}]_{1111}}
\newcommand{\wcqeEOT}{[\WC_{qe}]_{1211}}
\newcommand{\wcqeETT}{[\WC_{qe}]_{2211}}
\newcommand{\wcphidOT}{[\WC_{\phi d]_{12}}}
\newcommand{\wcphiqoneOO}{[\WC_{\varphi q}^{(1)}]_{11}}
\newcommand{\wcphiqoneOT}{[\WC_{\varphi q}^{(1)}]_{12}}
\newcommand{\wcphiqoneTT}{[\WC_{\varphi q}^{(1)}]_{22}}
\newcommand{\wcphiqthreeOO}{[\WC_{\varphi q}^{(3)}]_{11}}
\newcommand{\wcphiqthreeOT}{[\WC_{\varphi q}^{(3)}]_{12}}
\newcommand{\wcphiqthreeTT}{[\WC_{\varphi q}^{(3)}]_{22}}

\newcommand{\MET}{E{\!\!\!/}_T}

\begin{document}
\baselineskip=17pt \parskip=5pt

\title{High-$\rm{p_T}$ LHC constraints on SMEFT operators affecting rare kaon and hyperon decays}

\author{Arnab Roy}
\email{arnab.roy1@monash.edu}
\affiliation{School of Physics and Astronomy, Monash University, Wellington Road, Clayton, Victoria 3800, Australia}

\author{German Valencia}
\email{german.valencia@monash.edu}
\affiliation{School of Physics and Astronomy, Monash University, Wellington Road, Clayton, Victoria 3800, Australia}

\begin{abstract}

We consider the constraints on new physics affecting rare kaon and hyperon decay as parametrized by a low energy effective theory that can be obtained from high-$\rm p_T$ observables at LHC. To this end, we match the relevant low energy Wilson coefficients onto those in SMEFT at dimension six and compute their contributions to charged dilepton, monolepton, monojets, and $\rm VH$ searches by ATLAS and CMS. Constraints from rare kaon decays are generally more restrictive, with the ones arising from high-$\rm p_T$ measurements complementing them along directions in parameter space to which the former are not sensitive. We also illustrate the effect of a high luminosity LHC run in reducing the viable parameter regions.

\end{abstract}
\maketitle

\newpage

{\small\hypersetup{linkcolor=black}\tableofcontents}

\newpage

\section{Introduction}

The rare kaon decay program seeks precise measurements of standard model (SM) parameters and constraints on new physics (NP) through measurements of flavour-changing neutral current (FCNC) mediated modes \cite{Aebischer:2022vky,Anzivino:2023bhp}. Modes with neutrino pairs, such as $K\to \pi \nu\bar\nu$, can be calculated precisely \footnote{These calculations have been improved over many years, the latest improvement being the addition of two-loop electroweak corrections \cite{Brod:2010hi}.} and can lead to precise determinations of CKM matrix parameters. However, current experimental precision is insufficient for this purpose.  Modes with a charged dilepton pair suffer from considerable theoretical uncertainty, which weakens, but does not negate, the constraints on NP that can be obtained. 
When NP is parametrized using a low-energy effective theory (LEFT), rare kaon decays remain blind to certain directions in parameter space, which rare hyperon decays can partially cover \cite{He:2018yzu,Geng:2021fog,He:2023cqg,Roy:2024hqg}. 

The operators describing NP appearing in the LEFT may originate in scenarios beyond the SM.  One way to go beyond LEFT in a somewhat model-independent way is to match the LEFT onto an effective theory at the electroweak scale (SMEFT). This matching enables the inclusion of constraints arising from high-energy observables, such as those at LHC. Further, it relates the LEFT operators relevant for neutrino modes to those applicable for charged lepton modes. In this paper, we pursue this connection to obtain new constraints on the set of pertinent operators for rare kaon/hyperon decays related to the quark level process $s\to d \ell \bar\ell$ at dimension six. To this end, we consider those LHC observables sensitive to the parton level process $s d \to \ell\bar\ell$ for which measurements exist. In particular, we include dilepton, monolepton, and missing-$p_T$ (monojet) searches for NP from ATLAS and CMS.

Two recent studies \cite{Fajfer:2023nmz,Hiller:2024vtr} explored $|\Delta S|=1$ four-fermion operators using high-$\rm p_T$ observables. While \cite{Fajfer:2023nmz} focuses on dilepton measurements at the LHC, \cite{Hiller:2024vtr} also includes monojet searches to compare with low-energy observables. However, in comparing the effects of high-$\rm p_T$ and low-energy observables on the LEFT coefficients, these studies match the LEFT coefficients to 2Q2L-type SMEFT operators only, neglecting the contributions from 2Q2H, 2q2H, 2Q2e and 2q2e -type operators (where, `Q' and `L' are left-handed quark and lepton doublets respectively; `q' and `e' stand for the right-handed quark and lepton singlets, and H is the Higgs doublet; for notational simplicity we also use $Q^{\prime}\equiv \{Q,q\}$). Although the 2$Q^{\prime}$2H group of operators are mainly constrained by precision measurements at LEP\,\cite{ALEPH:2005ab,OPAL:2007ytu} and LHC, recent global fits (including top, Higgs, diboson, and electroweak observables)~\cite{Ellis:2020unq,Ethier:2021bye,Bartocci:2023nvp} suggest that the constraints on these 2$Q^{\prime}$2H-type operators are of the same order (and weaker in the case of marginalized fits) as the constraints on 2Q2L-type operators from monojet and Drell-Yan measurements \cite{Hiller:2024vtr,Bissmann:2020mfi}. On the other hand, monojet and Drell-Yan measurements constrain the 2$Q^{\prime}$2e-type operators at the same level as the 2Q2L ones~\cite{Hiller:2024vtr,Bissmann:2020mfi}. More than $\mathcal{O}(10\%)$ extra contribution is expected on the LEFT WCs from each of the 2$Q^{\prime}$2H and 2$Q^{\prime}$2e-type WCs compared to the 2Q2L ones. Therefore, as we enter the precision measurement era at the LHC,  analyses with the complete set of operators are necessary to assess `where we currently stand'. We pursue this goal in this work. Notably, in addition to Drell-Yan or monojet processes (which are weakly sensitive to the 2$Q^{\prime}$2H-type operators), we employ other measurements such as WH and ZH (VH) production, which also help in constraining the dipole operators, offering an additional benefit, and lead to tighter constraints.

The paper is organized as follows. In Section~\ref{sec:framework}, we outline the EFT framework and the relevant SMEFT-LEFT matching conditions. Section~\ref{sec:processes} describes our analysis setup for calculating various high-$\rm p_T$ LHC observables. In Section~\ref{sec:constraints}, we present the resulting constraints on the SMEFT Wilson coefficients (WCs), including high-luminosity projections. Section~\ref{sec:left_constraints} focuses on the constraints for the LEFT coefficients and relevant discussions regarding the complementarity of high-$\rm p_T$ and Kaon observables. In Section~\ref{sec:comparison}, we discuss key insights derived from the results in Sections~\ref{sec:constraints} and \ref{sec:left_constraints}. Finally, we summarise in Section~\ref{sec:summary}.

\section{Framework for new physics effects}
\label{sec:framework}
The effects of NP on rare kaon and hyperon decay can be parameterised with a low energy effective field theory (LEFT) with Hamiltonian, 
\begin{align}
{\cal H}_{\rm eff}^{} & \,=\, -\sum_i {\WC}_i{\cal O}_i+ {\rm h.c.} \,. \label{eq:npeff}
\end{align}
In this work we include all the dimension five and six operators ($\mathcal{O}_i$) that can affect the low energy kaon and hyperon decays to a final state containing a dilepton pair and that can be matched to a SMEFT operator up to dimension six \footnote{This excludes the tensor operators which only appear at dim-8 in SMEFT.}. We follow the notation of \cite{Roy:2024hqg}, which mostly matches that of the WET basis in {\tt flavio}\cite{Straub:2018kue}, except for not including an explicit factor of $m_s$ in the scalar (pseudoscalar) operators. We list explicitly the operators that contribute to rare hyperon decay, with their hermitian conjugates contributing to rare $K^+$ decay:
\begin{align}
{\cal O}_7^{} & = \frac{\lambda_t^{}G_{\rm F}^{}}{\sqrt2} \frac{{\tt e}\,m_s^{}}{4\pi^2}\, \overline{s_R^{}}\sigma^{\kappa\nu} d_L^{}\, F_{\kappa\nu}^{} \,, &
{\cal O}_{7}^\prime & = \frac{\lambda_t^{}G_{\rm F}^{}}{\sqrt2} \frac{{\tt e}\, m_s^{}}{4\pi^2}\, \overline{s_L^{}}\sigma^{\kappa\nu} d_R^{}\, F_{\kappa\nu}^{} \,, 
\nonumber \\
{\cal O}_9 & = \frac{\lambda_t^{}G_{\rm F}}{\sqrt2} \frac{\tt e^2}{4\pi^2}\, \overline{s_L^{}}\gamma^\nu d_L^{}\, \overline e_i\gamma_\nu^{}e_j \,, &
{\cal O}_{9}^\prime & = \frac{\lambda_t^{}G_{\rm F}^{}}{\sqrt2}  \frac{\tt e^2}{4\pi^2}\, \overline{s_R^{}}\gamma^\nu d_R^{}\, \overline e_i\gamma_\nu^{}e_j \,, \nonumber \\
{\cal O}_{10} & = \frac{\lambda_t^{}G_{\rm F}^{}}{\sqrt2} \frac{\tt e^2}{4\pi^2}\, \overline{s_L^{}}\gamma^\nu d_L^{}\, \overline e_i \gamma_\nu^{} \gamma_5^{}e_j \,, &
{\cal O}_{10}^\prime & = \frac{\lambda_t^{}G_{\rm F}^{}}{\sqrt2}  \frac{\tt e^2}{4\pi^2}\, \overline{s_R^{}}\gamma^\nu d_R^{}\, \overline e_i\gamma_\nu^{} \gamma_5^{}e_j \,, 
\nonumber \\
{\cal O}_S^{} & = \frac{\lambda_t^{}G_{\rm F}^{}}{\sqrt2} \frac{{\tt e}^2}{4\pi^2}\, \overline{s_L^{}} d_R^{}\, \overline e_ie_j \,, &
{\cal O}_{S}^\prime & = \frac{\lambda_t^{}G_{\rm F}^{}}{\sqrt2} \frac{{\tt e}^2}{4\pi^2}\, \overline{s_R^{}}d_L^{}\, \overline e_ie_j \,, 
\nonumber \\
{\cal O}_P^{} & = \frac{\lambda_t^{}G_{\rm F}^{}}{\sqrt2} \frac{{\tt e}^2}{4\pi^2}\, \overline{s_L^{}} d_R^{}\, \overline e_i \gamma_5^{}e_j \,, &
{\cal O}_{P}^\prime & = \frac{\lambda_t^{}G_{\rm F}^{}}{\sqrt2} \frac{{\tt e}^2}{4\pi^2} \overline{s_R^{}}d_L^{}\, \overline e_i\gamma_5^{}e_j \nonumber\\
{\cal O}_{L} & = \frac{\lambda_t^{}G_{\rm F}^{}}{\sqrt2} \frac{{\tt e}^2}{4\pi^2}\,\overline{s_L^{}}\gamma_\mu^{} d_L^{}\, \overline\nu_i \gamma^\mu (1-\gamma_5)\nu_j \,, &
    {\cal O}_{R} & = \frac{\lambda_t^{}G_{\rm F}^{}}{\sqrt2} \frac{{\tt e}^2}{4\pi^2}\,\overline{s_R^{}}\gamma_\mu^{} d_R^{}\, \overline\nu_i \gamma^\mu (1-\gamma_5)\nu_j \,. 
& \label{eq:LEFT_operators}
\end{align} 
In the above expressions, we use $\lambda_t=V_{ts}^\star V_{td}$, and the Wilson coefficients (WC), $\WC_k$ in Eq.\,\ref{eq:npeff}, refer only to possible NP contributions, with any SM contribution accounted for separately. The low-energy kaon and hyperon observables often select quark bilinears with definite parity, which makes it useful to refer to the combinations $\WC_{k\pm}\equiv\left(\WC_{k}\pm \WC_{k^\prime}\right)$. We will further limit ourselves to the case of CP conserving NP so that $\lambda_t\WC_k$ is assumed to be real. When lepton-flavour universality violating effects are present, we distinguish between muons and electrons with additional indices, i.e. $\WC_{k\pm}^{ij}$, where $i,j\in 1,2$ refer to the first two lepton generations. 


To derive bounds on the LEFT operators from LHC searches, we will match our low-energy effective theory into one that parametrizes physics beyond the SM at the electroweak scale, SMEFT. The SM lagrangian ($\mathcal{L}_{SM}$) is extended with  higher-dimensional operators up to dimension six as, 
\begin{align}
\mathcal{L}_{SMEFT} = \mathcal{L}_{SM}+\frac{1}{\Lambda^2}\sum_{i=1}^{N} C_i^{(6)}\mathcal{O}_i^{(6)}.
\label{eq:smeft}
\end{align}
The second term accounts for the NP contribution, comprising $\rm N$ dimension-6 operators ${\cal O}_i^{(6)}$ with associated Wilson coefficients $C_i^{(6)}$ and a cutoff scale $\Lambda$.
We will use the \textit{Warsaw basis}\,\cite{Grzadkowski:2010es}, where we find that the relevant operators up to dimension six that match onto Eq.~\ref{eq:LEFT_operators} are:
\begin{align}
\Op_{d W} &= \left( \bar q \sigma^{\mu \nu} d \right) \tau^I \varphi W_{\mu \nu}^I & 
\Op_{d B} &= \left( \bar q \sigma^{\mu \nu} d \right) \varphi B_{\mu \nu}\nonumber\\
\Op_{\ell q}^{(1)}  & =\left( \bar \ell \gamma_\mu \ell \right) \left( \bar q \gamma^\mu q \right), &
\Op_{\ell q}^{(3)}                &= \left( \bar \ell \gamma_\mu \tau^I \ell \right) \left( \bar q \gamma^\mu \tau^I q \right),\nonumber\\
\Op_{ed} &= \left( \bar e \gamma_\mu e \right) \left( \bar d \gamma^\mu d \right) & \Op_{\ell d} &= \left( \bar \ell \gamma_\mu \ell \right) \left( \bar d \gamma^\mu d \right), \nonumber\\
\Op_{\varphi q}^{(1)}&=\left( \varphi^\dagger i \overleftrightarrow{D}_\mu \varphi \right) \left( \bar q \gamma^\mu q \right) &
\Op_{\varphi q}^{(3)} &=\left( \varphi^\dagger i\tau^I \overleftrightarrow{D}_\mu \varphi \right) \left( \bar q \tau^I \gamma^\mu q \right) ,\nonumber\\
\Op_{q e} &= \left( \bar q \gamma_\mu q \right) \left( \bar e \gamma^\mu e \right) & \Op_{\varphi d}  &= \left( \varphi^\dagger i D_\mu \varphi \right) \left( \bar d \gamma^\mu d \right) ,\nonumber\\
\Op_{\ell edq} &= \left( \bar \ell e \right) \left( \bar d q \right).
\label{eq:SMEFT_operators}
\end{align}
Here, $q,\ell$ are the left-handed $\rm SU(2)_L$ doublets ($q = (u_L, d_L)^T, \ell = (\nu_L, e_L)^T$) ,  and $u,d,e$ are the right-handed singlets in the quark and lepton sectors respectively. The $\rm SU(2)_L$ and $\rm U(1)_Y$ gauge field strengths are denoted by $W_{\mu\nu}^{I}$ and $B_{\mu \nu}$, $\phi$ is the Higgs doublet, and $\overleftrightarrow{D}$ is the gauge covariant derivative acting on both sides, i.e., $( \varphi^\dagger i \overleftrightarrow{D}_\mu \varphi)=\phi^{\dagger}(iD_\mu\phi)-(iD_\mu \phi^{\dagger})\phi$, and all these operators carry flavour indices which we have omitted for notational simplicity.

Restricting our study to a CKM matrix that includes only Cabbibbo mixing, the matching conditions between the SMEFT and LEFT operators listed above at the electroweak scale ($\mu=\mu_{EW}$) are~\cite{Jenkins:2017jig,Alonso:2014csa}:

\begin{align}
    \WC_9^{ii} &= \frac{4\pi^2}{e^2\lambda_t}\frac{v^2}{\Lambda^2}\left(
    \left([\WC_{\ell q}^{(1)}] + [\WC_{\ell q}^{(3)}]\right)\otimes\left(
    [ii12](c_\theta^2-s_\theta^2)+(  [ii11]-[ii22])c_\theta s_\theta\right)\right.\nonumber\label{eq:matching1}\\
    &+[\WC_{q e}]\otimes\left(
    [12ii](c_\theta^2-s_\theta^2)+(  [11ii]-[22ii])c_\theta s_\theta\right)\nonumber\\
    &-\left.(1-4s_W^2)([\WC_{\varphi q}^{(1)}]+[\WC_{\varphi q}^{(3)}])\otimes\left(
    [12](c_\theta^2-s_\theta^2)+(  [11]-[22])c_\theta s_\theta\right)\right)\\
    \WC_9^{\prime ii} &= \frac{4\pi^2}{e^2\lambda_t}\frac{v^2}{\Lambda^2}\left([\WC_{\ell d}]_{ii12} + [\WC_{ed}]_{ii12} -(1-4s_W^2)[\WC_{\varphi d}]_{12}\right)\\
    \WC_{10}^{ii} &= \frac{4\pi^2}{e^2\lambda_t}\frac{v^2}{\Lambda^2}\left(
    [\WC_{q e}]\otimes\left(
    [12ii](c_\theta^2-s_\theta^2)+(  [11ii]-[22ii])c_\theta s_\theta\right)\right.\nonumber\\
    &\left.-\left([\WC_{\ell q}^{(1)}] + [\WC_{\ell q}^{(3)}]\right)\otimes\left(
    [ii12](c_\theta^2-s_\theta^2)+(  [ii11]-[ii22])c_\theta s_\theta\right)\right.\nonumber\\
    &\left.+([\WC_{\varphi q}^{(1)}]+[\WC_{\varphi q}^{(3)}])\otimes\left(
    [12](c_\theta^2-s_\theta^2)+(  [11]-[22])c_\theta s_\theta\right)\right)\label{eq:matching_C10}\\
    \WC_{10}^{\prime ii} &= \frac{4\pi^2}{e^2\lambda_t}\frac{v^2}{\Lambda^2}\left( [\WC_{ed}]_{ii12} -[\WC_{\ell d}]_{ii12} +[\WC_{\varphi d}]_{12}\right)\\
    \WC_{S}^{ii} &=  -\WC_{P}^{ii} = \frac{4\pi^2}{e^2\lambda_t}\frac{v^2}{\Lambda^2}[\WC_{\ell edq}^*]\otimes\left([ii12]c_\theta + [ii11]s_\theta\right)\\
    \WC_{S}^{\prime ii} &=  \WC_{P}^{\prime ii} = \frac{4\pi^2}{e^2\lambda_t}\frac{v^2}{\Lambda^2}[\WC_{\ell edq}]\otimes\left([ii21]c_\theta + [ii11]s_\theta\right)\\
    \WC_7 &= \frac{4\pi^2}{e m_s\lambda_t}\frac{v^2}{\Lambda^2}[\WC_{d\gamma}]\otimes\left([12]c_\theta + [11]s_\theta\right)\\
    \WC_7^{\prime} &= \frac{4\pi^2}{e m_s\lambda_t}\frac{v^2}{\Lambda^2}[\WC_{d\gamma}]\otimes\left([21]c_\theta + [11]s_\theta\right)\\
    \WC_L^{ii} &=\frac{4\pi^2}{e^2\lambda_t}\frac{v^2}{\Lambda^2}\left(
    \left([\WC_{\ell q}^{(1)}] - [\WC_{\ell q}^{(3)}]\right)\otimes\left(
    [ii12](c_\theta^2-s_\theta^2)+(  [ii11]-[ii22])c_\theta s_\theta\right)\right)\\
    \WC_R^{ii} &=\frac{4\pi^2}{e^2\lambda_t}\frac{v^2}{\Lambda^2}[\WC_{\ell d}]_{ii12},\label{eq:matching2}
\end{align}
where the index $i$ reflects the lepton generation, electron or muon or one of the three neutrino species. The notation $\otimes$ in Eq.~\ref{eq:matching1}-\ref{eq:matching2} simply denotes the specific combination of Cabbibbo angle that accompanies the different operators. For example, 
$\wclqone$ is short hand for the three possibilities in quark-flavour space $[\wclqone]_{ii12},~[\wclqone]_{ii11}$  and $[\wclqone]_{ii22}$. 

Note that both the SMEFT operators and the LEFT operators evolve under renormalization. We have checked the evolution of the SMEFT operators from $\mu=1$ TeV to $\mu=\mu_{EW}$ and of the LEFT operators from $\mu=\mu_{EW}$ to $\mu=1$ GeV using the \nolinkurl{DSixTools} package~\cite{Celis:2017hod,Fuentes-Martin:2020zaz}. We found the variations to be $\lesssim 10\%$ for all of them, below the uncertainty in our monojet and VH studies. For this reason, we do not include the effects of RG evolution on the WCs in our numerical study. 

The degeneracy of the $\WC_{S,P}^{(\prime)}$ LEFT coefficients appearing in Eq.~\ref{eq:matching2} is lifted at dimension eight in SMEFT. A complete study at dim-8 is beyond the scope of this work, but we will use one example that lifts this degeneracy, \cite{Murphy:2020rsh,Hamoudou:2022tdn}
\begin{align}
    \Op_{\ell e q d H^2}^{(3)} &=  \left( \bar \ell  e H\right) \left( \bar q  d H \right),
\end{align}
for illustration. Including this operator, the matching for (pseudo)-scalar operators becomes \cite{Hamoudou:2022tdn},
\begin{align}
	\WC_S^{\mu\mu}&=\frac{4\pi^2}{e^2\lambda_t}\dfrac{{v}^2}{\Lambda^2}\,\left([\WC_{\ell edq}^*]\otimes\left([2212]c_\theta + [2211]s_\theta\right)+\dfrac{{v}^2}{2\Lambda^2}\left[\WC_{\substack{leqdH^2}}^{(3)}\right]\otimes\left([2221]c_\theta + [2211]s_\theta\right)\right)\\
	\WC_P^{\mu\mu}&=\frac{4\pi^2}{e^2\lambda_t}\dfrac{{v}^2}{\Lambda^2}\,\left(-[\WC_{\ell edq}^*]\otimes\left([2212]c_\theta + [2211]s_\theta\right)+\dfrac{{v}^2}{2\Lambda^2}\left[\WC_{\substack{leqdH^2}}^{(3)}\right]\otimes\left([2221]c_\theta + [2211]s_\theta\right)\right)\\
 \WC_S^{\prime ^{\mu\mu}}&=\frac{4\pi^2}{e^2\lambda_t}\dfrac{{v}^2}{\Lambda^2}\,\left([\WC_{\ell edq}]\otimes\left([2221]c_\theta + [2211]s_\theta\right)+\dfrac{{v}^2}{2\Lambda^2}\left[\WC_{\substack{leqdH^2}}^{*(3)}\right]\otimes\left([2212]c_\theta + [2211]s_\theta\right)\right)\\
	\WC_P^{\prime ^{\mu\mu}}&=\frac{4\pi^2}{e^2\lambda_t}\dfrac{{v}^2}{\Lambda^2}\,\left([\WC_{\ell edq}]\otimes\left([2221]c_\theta + [2211]s_\theta\right)-\dfrac{{v}^2}{2\Lambda^2}\left[\WC_{\substack{leqdH^2}}^{*(3)}\right]\otimes\left([2212]c_\theta + [2211]s_\theta\right)\right)
 \label{dim8m}
\end{align}

 The high-$\rm p_T$ processes constraining the different operators in \ref{eq:SMEFT_operators} are summarized in Table~\ref{tab:LHC_searches}. In our study, we use the diagonal up-quark basis where the quark weak and mass eigenstates are related by
\begin{align}
    V_u=\mathbb{1},&\;\;\; V_d=V_{CKM}^\dagger. 
    \label{quarkbasis}
\end{align}
One consequence of the $\rm SU(2)_L\times U(1)_Y$ gauge invariance of SMEFT combined with the use of the diagonal up-quark basis is reflected in the specific contributions from up-type partons as well as flavour diagonal $q\bar{q}$ partons initiating the high-$\rm p_T$ processes as summarized in Table~\ref{tab:LHC_searches}.

As the Table suggests, there are as many as 24 SMEFT operators that contribute in some form to the processes we study. Their Wilson coefficients can be constrained by different LHC measurements as schematically shown on the Table.
  \begin{table}[]
 	\centering
 	\begin{tabular}{ c|c||c|c|c|c|c|c|c|c|c|c|c}
 		\hline
        \hline
 		$p p\downarrow$ & partons& $\opdw$ & $\opdb$ & $\oplqone$ &$\oplqthree$ & $\oped$ & $\opld$ & $\opphiqone$ & $\opphiqthree$ & $\opqe$ & $\opphid$ & $\opledq$ \\
 		\hline
        \hline
 		\multirow{3}{*}{$\ell\ell$}& $\{ds\}_{12},\{dd\}_{11},\{ss\}_{22}$ & $\checkmark$ & $\checkmark$ & $\checkmark$ & $\checkmark$ & $\checkmark$ & $\checkmark$ & $\checkmark$ & $\checkmark$ & $\checkmark$ & $\checkmark$ & $\checkmark$\\ 
 		& $\{dd,ss\}_{12},\{ds,sd\}_{11,22}$ & $\checkmark$ & $\checkmark$ & $\checkmark$  & $\checkmark$ & $\checkmark$ & $\checkmark$ & $\checkmark$ & $\checkmark$ & $\checkmark$ & $\checkmark$ & $\checkmark$\\ 
 		& $\{uc\}_{12},\{uu\}_{11},\{cc\}_{22}$ & - & - & $\checkmark$ & $\checkmark$ & - & - & $\checkmark$ & $\checkmark$ & $\checkmark$ & - & -\\ \hline
 		\multirow{3}{*}{$\ell\nu$} & $\{us,dc,du,cs\}_{12}$ & $\checkmark$ & - & - & $\checkmark$ & - & - & - & $\checkmark$ & - & - &$\checkmark$\\ 
 		& $\{ud,us\}_{11}$ & $\checkmark$ & - & - & $\checkmark$ & - & - & - & $\checkmark$ & - & - &$\checkmark$\\ 
 		& $\{cs,cd\}_{22}$ & $\checkmark$ & - & - & $\checkmark$ & - & - & - & $\checkmark$ & - & - &$\checkmark$\\ \hline
 		\multirow{3}{*}{WH}& $\{us,dc,du,cs\}_{12}$ & $\checkmark$ & - & - & - & - & - & - & $\checkmark$ & - & - &-\\ 
 		& $\{ud,us\}_{11}$ & $\checkmark$ & - & - & - & - & - & - & $\checkmark$ & - & - &-\\
 		& $\{cs,cd\}_{22}$ & $\checkmark$ & - & - & - & - & - & - & $\checkmark$ & - & - &-\\ \hline
 		\multirow{3}{*}{ZH}& $\{ds\}_{12}$ , $\{dd\}_{11}$, $\{ss\}_{22}$& $\checkmark$ & $\checkmark$ & - & - & - & - & $\checkmark$ & $\checkmark$ & - & $\checkmark$ & -\\
 		& $\{dd,ss\}_{12}, \{ds,sd\}_{11,22}$ & $\checkmark$ & $\checkmark$ & - & - & - & - & $\checkmark$ & $\checkmark$ & - & $\checkmark$ &-\\
 		& $\{uc\}_{12},\{uu\}_{11},\{cc\}_{22}$ & - & - & - & - & - & - & $\checkmark$ & $\checkmark$ & - & - & -\\ \hline
 		\multirow{3}{*}{$j\nu\nu$}& $\{ds\}_{12}$ , $\{dd\}_{11}$, $\{ss\}_{22}$ & $\checkmark$ & $\checkmark$ & $\checkmark$ & $\checkmark$ & - & $\checkmark$ & $\checkmark$ & $\checkmark$ & - & $\checkmark$ & - \\
 		& $\{dd,ss\}_{12}, \{ds,sd\}_{11,22}$  & $\checkmark$ & $\checkmark$ & $\checkmark$ & $\checkmark$ & - & $\checkmark$ & $\checkmark$ & $\checkmark$ & - & $\checkmark$ &-\\
 		& $\{uc\}_{12},\{uu\}_{11},\{cc\}_{22}$ & - & - & $\checkmark$ & $\checkmark$ & - & - & $\checkmark$ & $\checkmark$ & - & - & -\\
 		\hline
        \hline
 	\end{tabular}
 	\caption{\small LHC searches relevant to the SMEFT operators considered in this study. All operators have an implicit flavor structure involving first- and second-generation quarks and leptons, where applicable. The second column lists possible initial states for different quark-flavour structures of the SMEFT operators, denoted as subscripts to the brackets. One of the two fields for each pair inside the brackets is implicitly a conjugate.}
 	\label{tab:LHC_searches}
 \end{table}

\section{High-$\rm p_T$ observables at LHC}
\label{sec:processes}

In this section, we establish the setup to constrain the SMEFT operators in Eq.\,\ref{eq:SMEFT_operators} from the most sensitive current measurements, $\mathcal{L}\simeq 140\:fb^{-1}$ at the LHC. While most of the operators are dominantly sensitive to the $pp\to \ell\ell,\ell\nu$ searches, $\Op_{\varphi q}^{(1)}$, $\Op_{\varphi q}^{(3)}$, and $\Op_{\phi d}$ are not. Instead, these are sensitive to Higgs production in association with vector bosons (VH). Monojet searches are also affected by several of the operators. Table\;\ref{tab:LHC_searches} provides a guide as to which processes are sensitive to which operators.  The second column of this table represents the possible initial states for different operators, where a subscript (11, 22, or 12) indicates the quark-flavor structure of a given operator. For example, a checkmark under $\Op_{dW}$ for $\{ds\}_{12}$ implies that $[\Op_{dW}]_{12}\equiv\left( \bar d \sigma^{\mu \nu} s \right) \tau^I \varphi W_{\mu \nu}^I$ is sensitive to the $ds$ initial state. It should be apparent from this table that flavor-conserving quark bilinears, as well as up-quark type bilinears, will contribute as much, if not more, than their corresponding $|\Delta S|=1$ counterparts to the resulting constraints. While the specifics of which non $|\Delta S|=1$  contributions enter the results depend on the choice in Eq.~\ref{quarkbasis}, there will be analogous contributions for other choices. Our choice minimizes the up-quark contributions, which will always involve at least one charm PDF, whereas the choice of a down-diagonal basis adopted in \cite{Hiller:2024vtr} removes flavor diagonal down-type quark bilinear contributions.

\subsection{Constraints from VH searches}

For WH and ZH processes, we use the ATLAS measurement in Ref.~\cite{ATLAS:2020jwz}, where the cross-sections are measured and reported in four simplified template cross section (STXS) regions (Table~6 of \;\cite{ATLAS:2020jwz}), which we reproduce as  Table~\ref{tab:ATLAS_VH} in Appendix~\ref{appvh} for convenience. From these results, we extract the window for new physics that will be used to constrain the WC as $\sigma^{NP}=\sigma^{exp}-\sigma^{SM}$. 

We then calculate the fiducial cross-sections of WH and ZH within the SMEFT as a function of the relevant SMEFT WC (Eq.~\ref{eq:SMEFT_operators}), using the \nolinkurl{SMEFTsim_general_MwScheme_UFO} package \cite{Brivio:2017btx,Brivio:2020onw},  which is a \nolinkurl{Feynrules}~\cite{Alloul:2013bka} implementation of SMEFT at leading order (LO) in the Warsaw basis. We use \nolinkurl{MadGraph5-aMC@NLO-v3.5.4}\,\cite{Alwall:2014hca} to calculate the fiducial cross-sections of WH and ZH in the four STXS regions as quadratic polynomials in the WCs,
\begin{align}
    \sigma^{NP,MG5}=
\sum_{i=1}^{n}a_i C_i+\sum_{j\leq k}^{n}b_{jk}C_jC_k.
\label{eq:mu}
\end{align}
The numerical values of the coefficients ($a_i, b_{jk}$) are determined by fitting the cross-sections, treating the WCs as variables and $a_i, b_{jk}$ as fitting parameters (see Ref.~\cite{Guchait:2022ktz} for a detailed discussion of this method).
Explicit values for these coefficients are relegated to Appendix~\ref{appvh}. Identifying the RHS of Eq.~\ref{eq:mu} with the NP window $\sigma^{exp}-\sigma^{SM}$ defined above we calculate the $\chi^2$ as,
\begin{align}
    \chi^2(VH)=\sum_i \frac{\left(\sigma_{VH}^{exp}(i)-\sigma_{VH}^{SM}(i) - \sigma_{VH}^{MG5}(i)\right)^2}{\left[\delta\sigma_{VH}^{exp}(i)\right]^2},
\end{align}
where the sum runs over the four STXS regions. Since the theory uncertainties are significantly smaller than the quoted total experimental uncertainty, we ignore them. Our results in Eq.~\ref{WHpoly} indicate that the dominant EFT effect for the WH mode comes from the $\opphiqthree$, with a subleading contribution from the $\opdw$. Other operators do not contribute to WH, whereas ZH is also sensitive to $\opphiqone$, $\opphid$, and $\opdb$.

\subsection{Constraints from mono-jet searches}

The semileptonic SMEFT operators, $\oplqone,\oplqthree,\opld$, as well as the semi-bosonic operators, $\opdw,\opdb,\opphiqone,\opphiqthree,\opphid$ affect monojet production through the process $p p \to \nu\nu+j$ and their coefficients will therefore be constrained by monojet measurements at the LHC. Ref.~\cite{Hiller:2024vtr} recently investigated these constraints using the ATLAS measurement from Ref.~\cite{ATLAS:2021kxv}. In this work, we repeat the recast of the ATLAS analysis to include it in our \textit{combined} $\chi^2$ from all processes. The relevant results of Ref.~\cite{ATLAS:2021kxv} are reproduced in Appendix \ref{appmj} for convenience.

For our recast we use \nolinkurl{MadGraph5-aMC@NLO-v3.5.4}\,\cite{Alwall:2014hca} with a PDFset \nolinkurl{PDF4LHC15_nlo_mc}\,\cite{Butterworth:2015oua} to simulate events with the \nolinkurl{SMEFTsim_general_MwScheme_UFO}\,\cite{Brivio:2017btx,Brivio:2020onw}.  Showering and hadronization are performed using \nolinkurl{Pythia8}\,\cite{Sjostrand:2006za,Sjostrand:2007gs} for all the events and \nolinkurl{Delphes3}\,\cite{deFavereau:2013fsa} is used to take into account the detector effects using the ATLAS detector card, including pile-up. This detector card applies all detector efficiencies and resolutions and also constructs physical objects, such as jets, leptons, and missing-$p_T$. Jets are refined with pile-up subtraction. Following Ref.~\cite{ATLAS:2021kxv} we use the Anti-KT jet reconstruction algorithm with a radius parameter 0.4. The efficiencies of other objects are also modified to match Ref.~\cite{ATLAS:2021kxv}. Once we have all the objects from \nolinkurl{Delphes3}, we perform the analysis with the same event selection criteria as Ref.~\cite{ATLAS:2021kxv}. To validate our recast, we generate the dominant SM backgrounds and check the event yields in different $\rm \MET$ bins, finding that they match the ATLAS estimation of the SM backgrounds to within $\sim 7-20\%$. This slight mismatch between the two estimations is expected, as our approach does not precisely account for all sources of uncertainties included in the partially data-driven experimental analysis. However, the impact of this mismatch on the calculation of the effects of different WCs is negligible compared to the uncertainty in experimental event yields.
 
With this MC setup,  we proceed to compute the number of monojet events ($\mathcal{N}=\sigma\times\mathcal{L}\times\epsilon$) as polynomials in the WCs for each $\rm \MET$-bin:
\begin{align}
\mathcal{N}^{NP}(\vec{\WC})=\mathcal{N}^{EFT}(\vec{\WC})-\mathcal{N}^{SM}=\sum_{i=1}^{n} \WC_i\beta_i+\sum_{j\leq k}^{n}\WC_j\WC_k\gamma_{jk},
\label{eq:N}    
\end{align}
with explicit coefficients tabulated in Appendix~\ref{appmj}. Note that for the dipole operators, only the combination $\WC_{dZ}=\cos{\theta_W}\,\wcdw+\sin{\theta_W}\,\wcdb$ appears in this case. Our results suggest that these events are mostly sensitive to $\WC_{dZ}$, followed by $\wcphiqone$, $\wcphiqthree$, $\wclqone$ and $\wclqthree$, where only the later four interfere with SM.

We then use these numbers in combination with the ATLAS results to construct a $\chi^2$. The NP window, $\Delta\mathcal{N}^{exp}$ is extracted by subtracting from the observed number of events for each $\rm \MET$ bin, the SM backgrounds reported by ATLAS. Consequently, we write the $\chi^2$ as:
\begin{align}
    \chi^2(MJ)=\frac{\left(\Delta\mathcal{N}^{exp} - \mathcal{N}^{NP}(\WC)\right)^2}{\left[\delta\left(\Delta\mathcal{N}\right)\right]^2}.
\end{align}
The uncertainty $\delta\left(\Delta\mathcal{N}\right)$ appearing in this formula is dominated by that of the ATLAS SM background extraction.

\subsection{Constraints from dilepton and monolepton searches}

The use of these two processes to constrain SMEFT operators is greatly simplified by the use of the package \nolinkurl{HighPT}~\cite{Allwicher:2022mcg,Allwicher:2022gkm}, which directly gives us the corresponding $\chi^2$. The expressions for this $\chi^2$ are too lengthy to reproduce here but are available from the authors upon request.  These processes are the only ones sensitive to $\opledq$, and the dilepton process is the only one sensitive to  $\oped$, and $\opqe$. As we will see, a comparison of constraints obtained from these two processes will allow us to differentiate between dim-6 contributions to scalar LEFT operators and a particular dim-8 operator that splits the degeneracy between them, as discussed above.

\section{Constraints on SMEFT coefficients}
\label{sec:constraints}

\begin{figure}
	\centering
	\includegraphics[width= 4.8 cm]{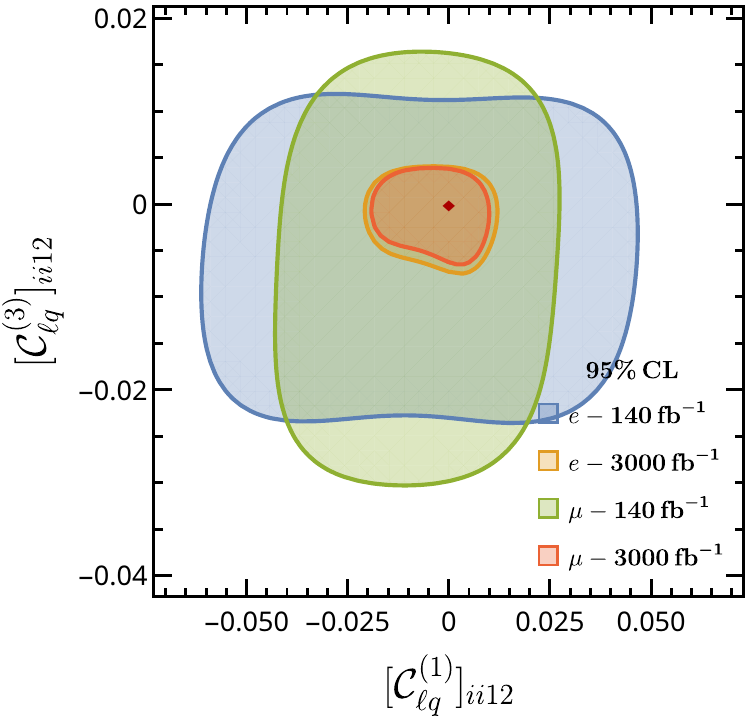}
	\includegraphics[width=4.8 cm]{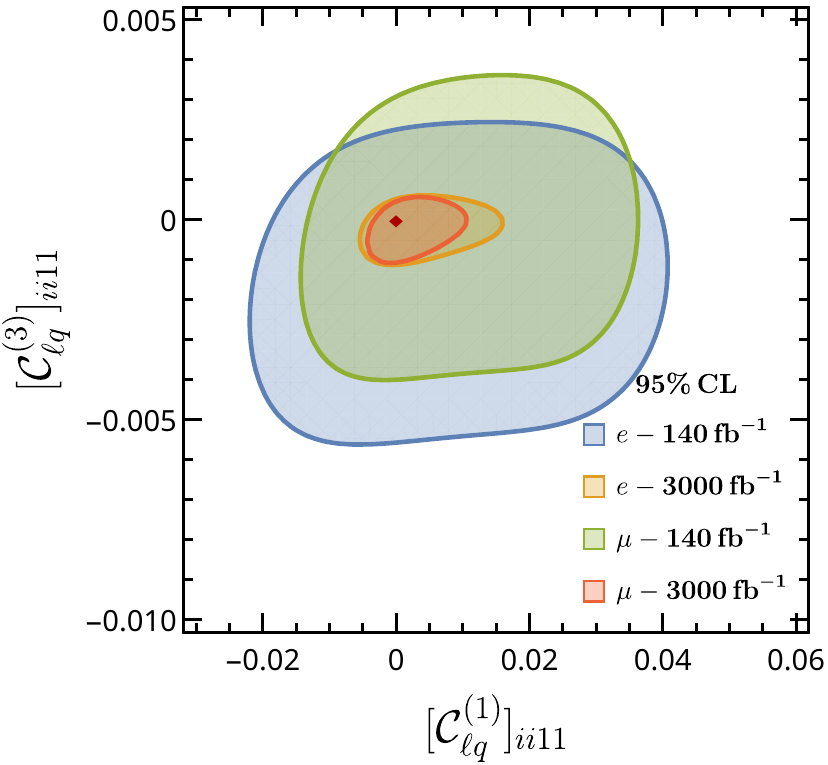}
		\includegraphics[width=4.8 cm]{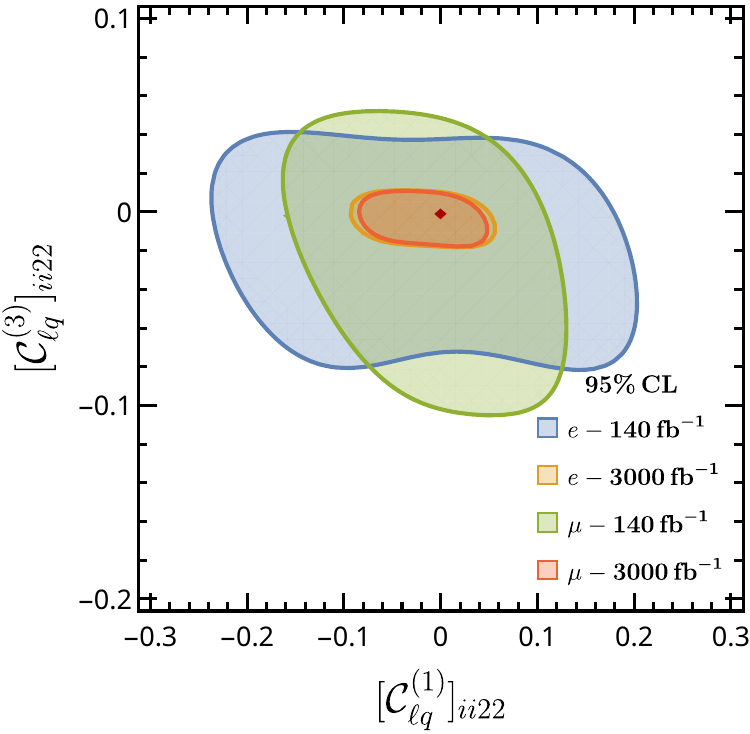}
	\caption{\small Two-dimensional slices of the combined 95$\%$ CL allowed (and projected) regions taken through the minimum for the SMEFT WCs \{$[\wclqone]_{ii12},[\wclqthree]_{ii12}$\} (left), \{$[\wclqone]_{ii11},[\wclqthree]_{ii11}$\} (middle), and \{$[\wclqone]_{ii22},[\wclqthree]_{ii22}$\} (right) for $\mathcal{L}=140\;\rm fb^{-1}$ (blue, green), and $\mathcal{L}=3000\;\rm fb^{-1}$ (yellow, red). Results are presented for both the electron ($i=1$) and muon ($i=2$) related operators.}
	\label{fig:smeft_95CL1}
\end{figure}

\begin{figure}
	\centering
	\includegraphics[width=4.8 cm]{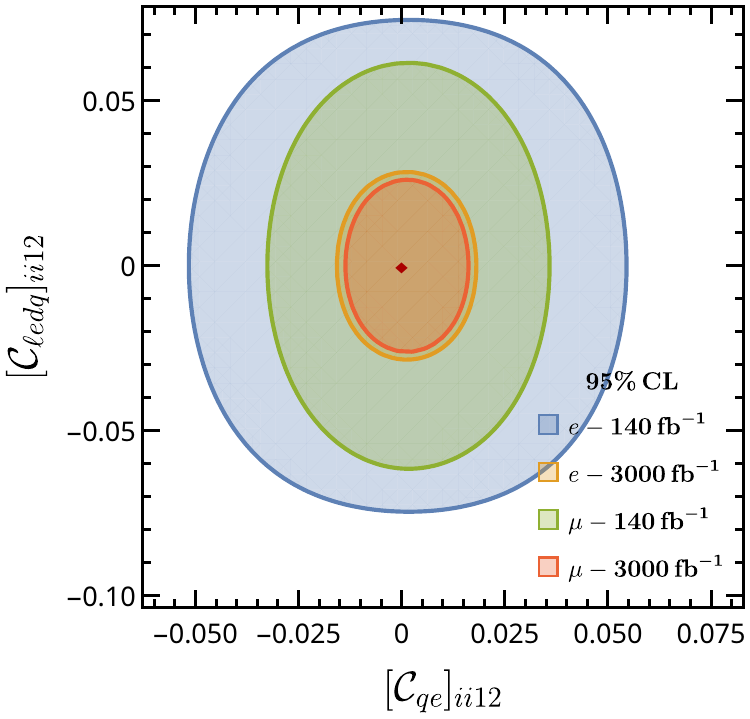}
	\includegraphics[width= 4.8 cm]{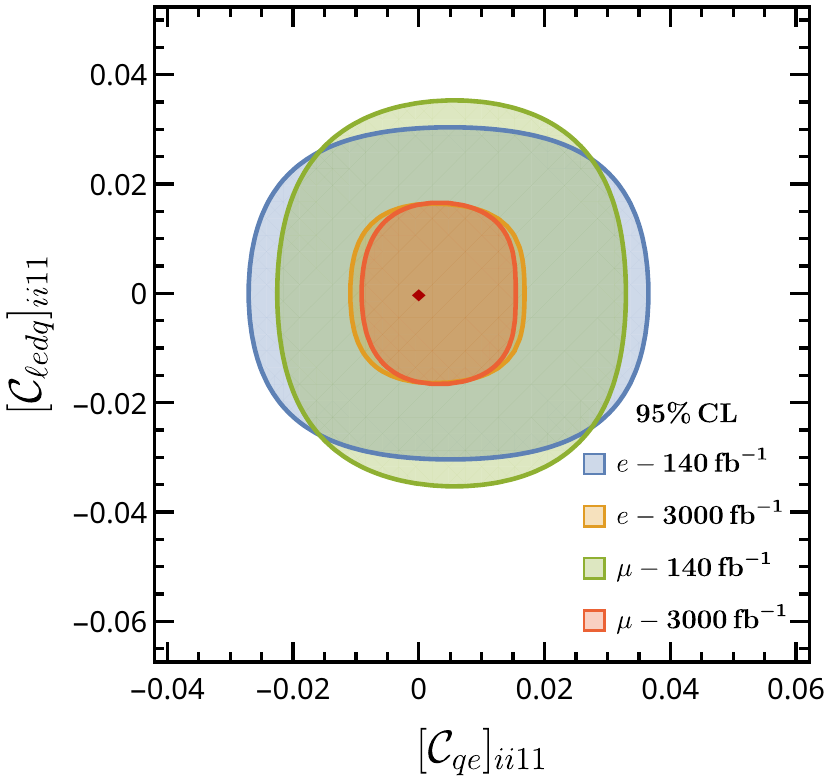}
	\includegraphics[width= 4.8 cm]{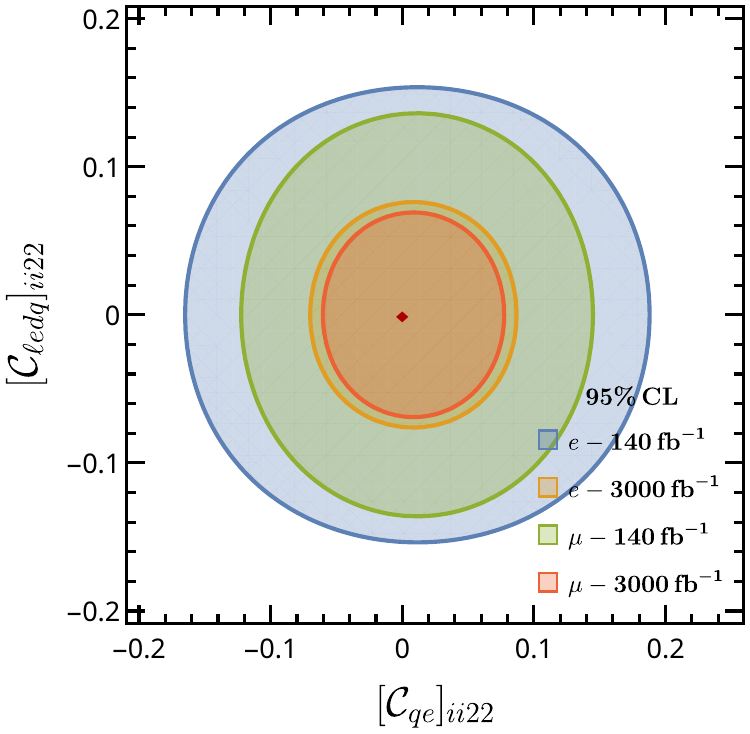}
	\caption{\small Same as Fig.~\ref{fig:smeft_95CL1} but for  \{$[\wcqe]_{ii12},[\wcledq]_{ii12}$\} (left), \{$[\wcqe]_{ii11},[\wcledq]_{ii11}$\} (middle), and \{$[\wcqe]_{ii22},[\wcledq]_{ii22}$\} (right).}
	\label{fig:smeft_95CL2}
\end{figure}

To constrain the SMEFT coefficients we use the combined $\chi^2$ for the different processes discussed above as,
\begin{align}
    \chi^2_{comb}=\chi^2_{\ell\ell}+\chi^2_{\ell\nu}+\chi^2_{VH}+\chi^2_{monojet}.
\end{align}
 We minimize this $\chi^2_{comb}$ for a total 24 degrees of freedom and find the $95\%$ CL allowed ranges of the SMEFT operators\footnote{Note that while considering the 12-combination, we minimize only the relevant 11 degrees of freedom, setting others to zero.}. We present our results as two-dimensional slices of this 24-dimensional 95$\%$ CL region, where the slice is taken through the minimum of the $\chi^2$ function. Fig.\;\ref{fig:smeft_95CL1}, \ref{fig:smeft_95CL2}, and \ref{fig:smeft_95CL4} (left) illustrates the results for the 2Q2L-type operators. The blue (electron) and green (muon) regions show the two-dimensional slices of the $95\%$ CL allowed ranges compatible with the RUN-II LHC measurements ($\mathcal{L}=140\; \mathrm{fb^{-1}}$). Whereas, the yellow (electron) and red (muon) regions show the projection for $\mathcal{L}=3000\; \mathrm{fb^{-1}}$, which is obtained by rescaling the number of events with the luminosity, i.e., $\mathcal{N}\to \mathcal{N}\times\left(3000/140\right)$, and the background uncertainty by $\sqrt{3000/140}$.  
All these semileptonic four-fermion operators are strongly constrained, roughly in the range $-0.05\lesssim\,\WC_{4F}\lesssim 0.05$, as seen in   Fig.\;\ref{fig:smeft_95CL1}  because of the precise dilepton and monolepton measurements. 

\begin{figure}
	\centering
	\includegraphics[width=4.8 cm]{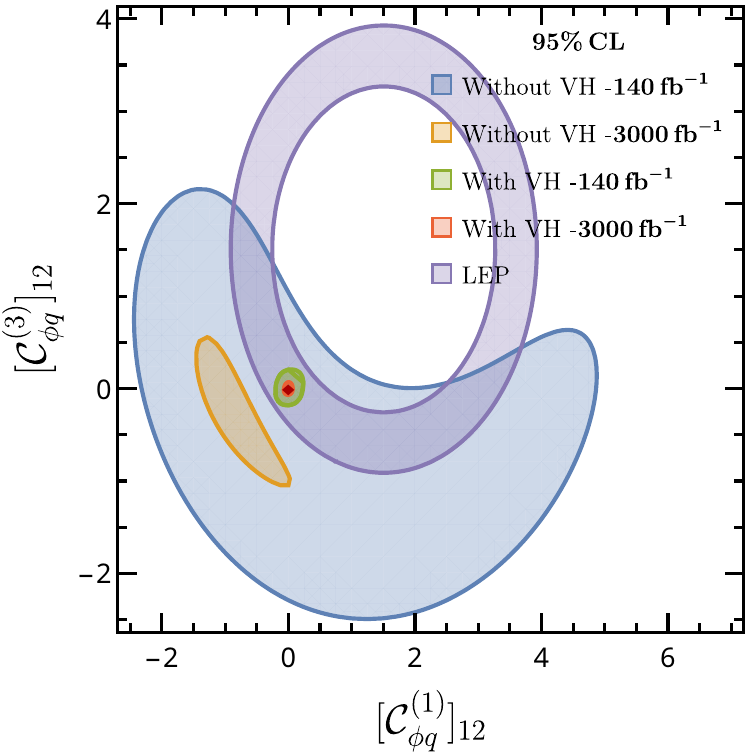}
	\includegraphics[width= 5 cm]{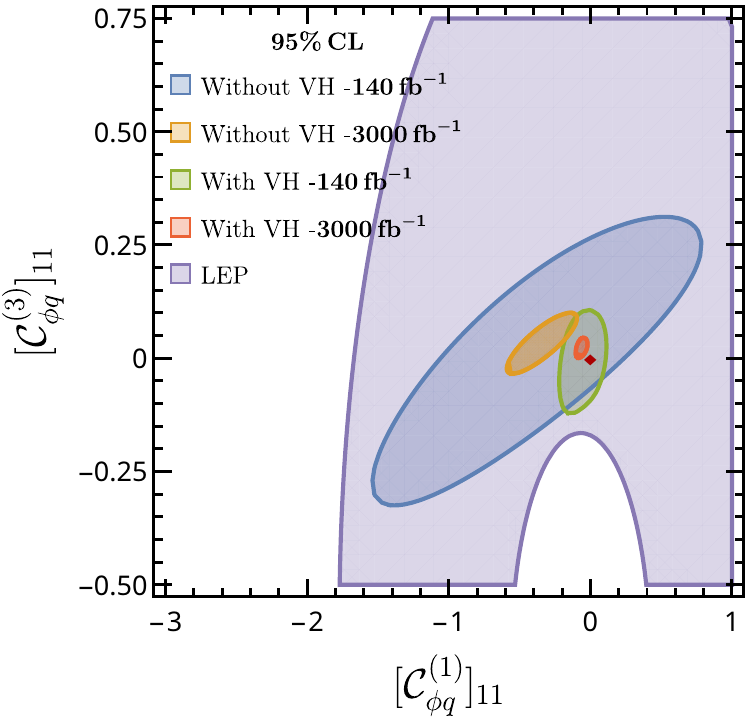}
	\includegraphics[width= 4.8 cm]{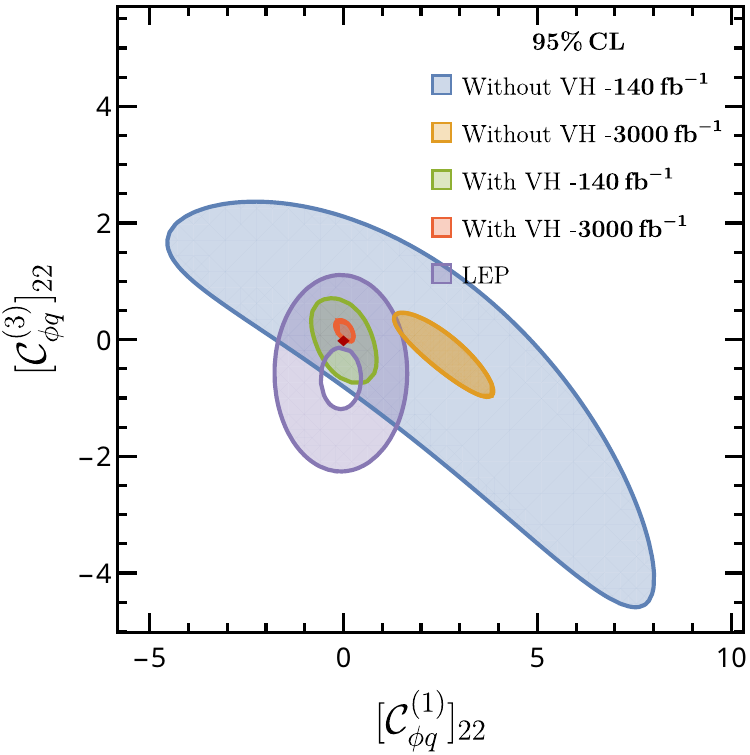}
	\caption{\small Two-dimensional slices of the combined 95$\%$ CL allowed (and projected) regions taken through the minimum \textit{considering the effects of VH measurements (green, red) or excluding them (blue, yellow)} for the SMEFT WCs  \{$[\wcphiqone]_{12},[\wcphiqthree]_{12}$\} (left), \{$[\wcphiqone]_{11},[\wcphiqthree]_{11}$\} (middle), and \{$[\wcphiqone]_{22},[\wcphiqthree]_{22}$\} (right). The luminosity options are $\mathcal{L}=140\;\rm fb^{-1}$ (blue, green), and $\mathcal{L}=3000\;\rm fb^{-1}$ (yellow, red). The indigo regions present the constraints from Z and W-width measurements at LEP.}
	\label{fig:smeft_95CL3}
\end{figure}

\begin{figure}
    \centering
    \includegraphics[width=6.5 cm]{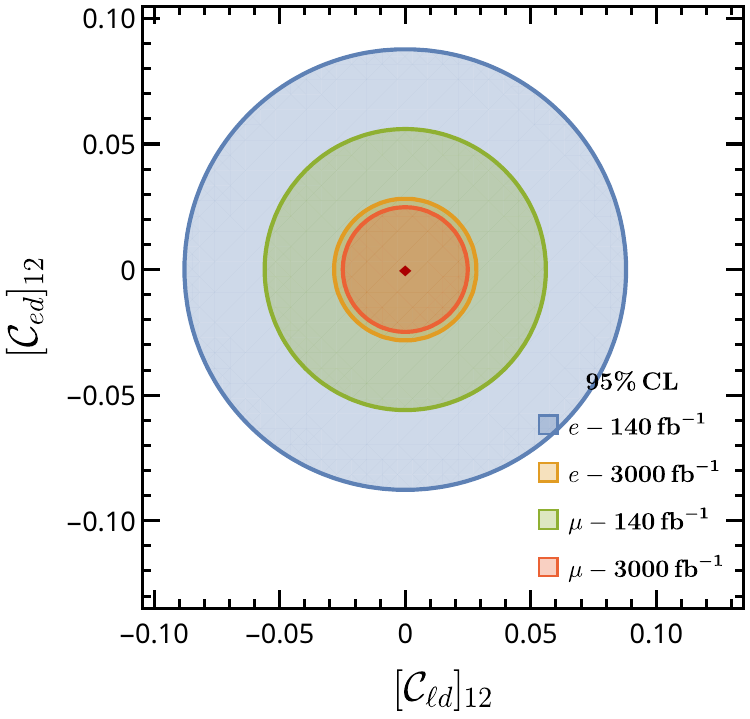}
    \includegraphics[width= 6 cm]{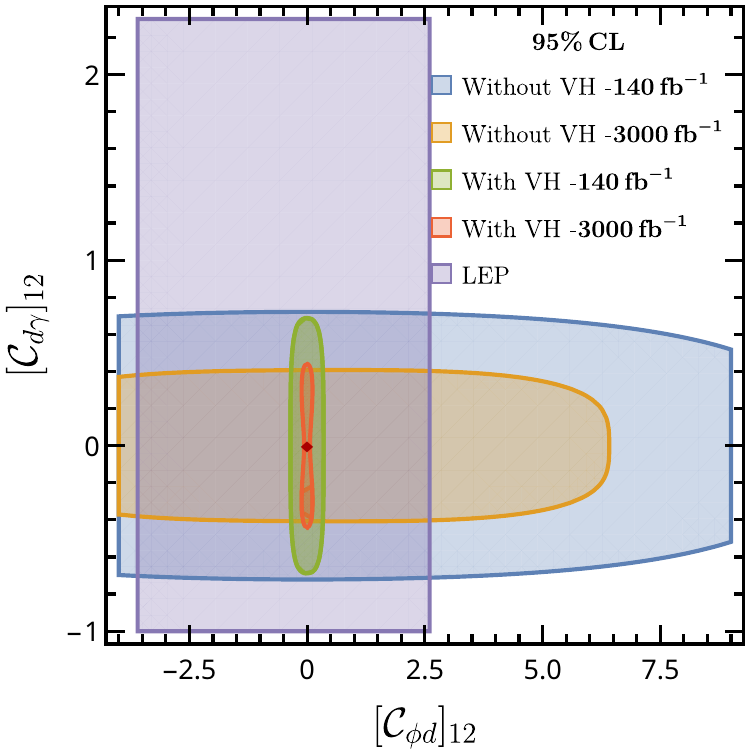}
    \caption{\small (Left) Same as Fig.~\ref{fig:smeft_95CL1} but for the SMEFT WCs \{$[\wcphiqone]_{12},[\wcphiqthree]_{12}$\}; (Right) Same as Fig.~\ref{fig:smeft_95CL3} but for the operators \{$[\wcphid]_{12},[\wcdgamma]_{12}$\}.}
    \label{fig:smeft_95CL4}
\end{figure}

To constrain the 2$Q^{\prime}$2H-type operators, the WH and ZH searches (VH) are crucial. These results are shown in Fig.~\ref{fig:smeft_95CL3}  and Fig.~\ref{fig:smeft_95CL4} (right).  In these figures, the blue and yellow regions represent the allowed parameter space {\it without} including VH processes (i.e., $\ell\ell+\ell\nu+$Monojet) for current and projected luminosities, while the green and red regions show the corresponding constraints when VH measurements are included. The bounds on the corresponding WCs are almost an order of magnitude looser than those for 4F operators: roughly in the range $-0.3\lesssim \WC_{Q^{\prime}2H}\lesssim 0.3$. These 2$Q^{\prime}$2H-type operators are independent of lepton flavor; however, the corresponding high-${\rm p_T}$ observables are affected by different experimental issues resulting in similar but not identical constraints as seen in Table \ref{tab:individual_SMEFT}. For comparison, we also show the bounds on these $\rm 2Q^{\prime}2H$ operators from  LEP+SLC measurement of the Z width and the LEP+CDF+D0 measurement of the W width, collectively denoted as ‘LEP’ for short (Fig.~\ref{fig:smeft_95CL3} and \ref{fig:smeft_95CL4} (right)).  The relevant calculations are presented in Appendix~\ref{app:Width_calculations}. The bounds from the LHC, including VH measurements, are much stronger than those from LEP. Therefore, we rely solely on high-$\rm p_T$ LHC observables to derive combined constraints on the LEFT operators.

At the SMEFT level, the difference in constraints seen in Figures~\ref{fig:smeft_95CL1}-\ref{fig:smeft_95CL3} is roughly due to a combination of experimental sensitivity to different processes, the contribution to each process by a given operator, and the PDFs involved. The different figures illustrate the importance of different processes and operators, while the distinction between flavour indices `12', `11', and `22' reflects the impact of the larger parton distribution functions (PDFs) of the up and down quarks compared to those of the strange and charm quarks (see Table \ref{tab:LHC_searches} for a guide to the initial partons associated with different operators). Looking at the  2Q2L operators, for example, the diagonal 11-flavour combination is the most strongly constrained by about a factor of four, followed by 12-flavour, while the 22-flavour combination is the weakest. This hierarchy is the same for the $\WC_{\ell edq}$ and $\WC_{qe}$ operators although the relative size between different flavour combinations is not quite the same. For the $\rm 2Q^{\prime}2H$ operators on the other hand, the best constraints arise from $VH$ measurements as can be seen in Figure~\ref{fig:smeft_95CL3}. For this observable, the hierarchy among flavor combinations remains the same, but with different relative sizes (see Table~\ref{tab:individual_SMEFT} for an overview of these relative sizes). All of these factors will play a role in the translation of these bounds to the LEFT coefficients as we discuss below.

\subsection{Uncertainties in high-luminosity projections} 
\label{sec:uncertainties}

\begin{figure}
    \centering
    \includegraphics[width= 5.5 cm]{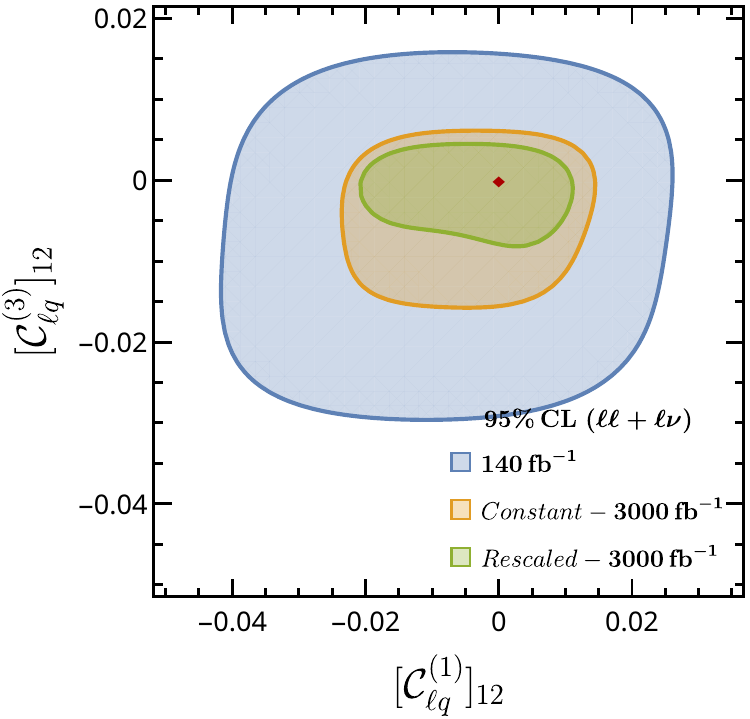}
    \includegraphics[width=5.5 cm]{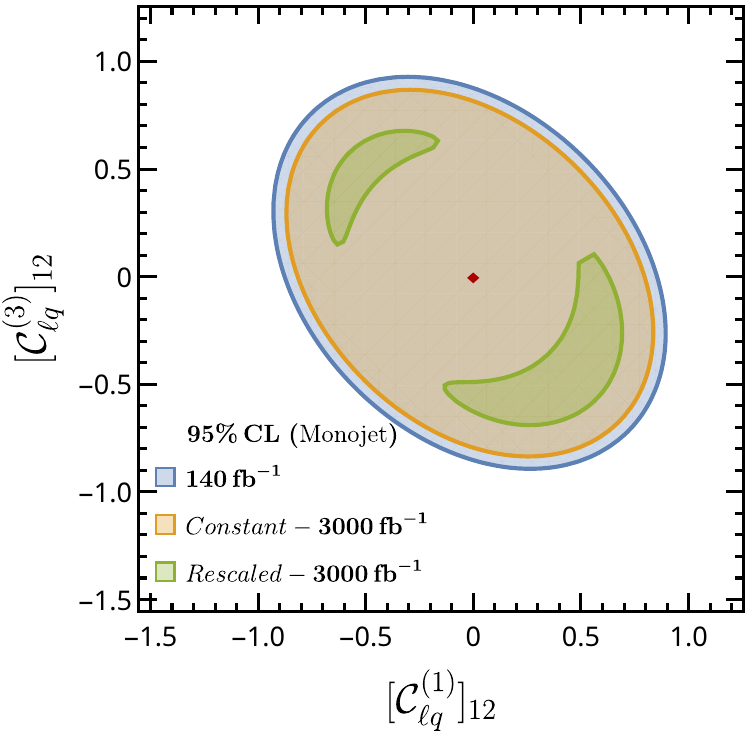}
    \includegraphics[width= 5.5 cm]{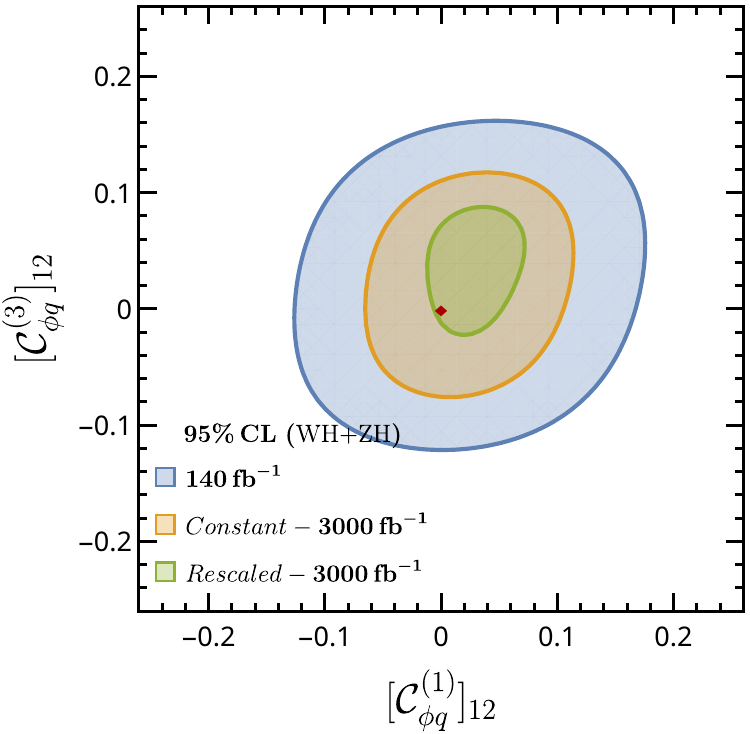}
    \caption{\small Effects of different methods for treating background uncertainty on the 95$\%$ CL regions for various SMEFT WCs, using $\ell\ell+\ell\nu$ (left), monojet (middle), and WH+ZH (right) measurements at $\mathcal{L}=140\;\rm fb^{-1}$ (blue) and $3000\;\rm fb^{-1}$ (yellow, green). The statistical uncertainty $\delta_{stat} = \sqrt{\mathcal{N}^{obs}}$ is always scaled as $\sqrt{3000/140}$. The systematic uncertainty is treated in two ways: (a) rescaled, where $\delta_{sys} \to \sqrt{3000/140} , \delta_{sys}$ (green), and (b) constant, where $\delta_{sys} \to (3000/140) , \delta_{sys}$ (yellow).}
    \label{fig:smeft_95CLunc}
\end{figure}

The treatment of uncertainties plays a crucial role in high-luminosity projections at the LHC. While the statistical uncertainty, $\delta_{stat} = \sqrt{\mathcal{N}^{obs}}$, scales as $\sqrt{3000/140}$, projections for the background uncertainty can be treated in two ways. An optimistic approach assumes an improvement in systematic uncertainty to match the increased statistics, $\delta_{sys} \to \sqrt{3000/140}\,\delta_{sys}$ (rescaled). In contrast, a conservative approach assumes the ratio of background uncertainty to background yield remains unchanged: $\delta_{sys} \to (3000/140)\,\delta_{sys}$ (constant). In Fig.~\ref{fig:smeft_95CLunc}, we show the impact of these two approaches on the high-luminosity projections for the $\ell\ell+\ell\nu$, monojet, and VH processes separately. The blue region represents the allowed 95$\%$ CL region at $\mathcal{L}=140\;\rm fb^{-1}$, while the yellow and green regions show the projections at $\mathcal{L}=3000\;\rm fb^{-1}$ for the ``constant'' and ``rescaled'' uncertainty treatments, respectively. For the ``rescaled'' background uncertainty, the expected deviation of the observed monojet events from the SM background at $\mathcal{L}=3000\;\rm fb^{-1}$ exceeds the uncertainty~\cite{ATLAS:2021kxv}, which is why the green region for the monojet measurement does not enclose the SM (red point) in Fig.~\ref{fig:smeft_95CLunc} (middle), and yellow region in Fig.~\ref{fig:smeft_95CL3}. In this study, we use the ``rescaled'' background uncertainty in all the plots, unless stated otherwise.

\section{Constraints on LEFT operators: Complementarity with rare kaon and hyperon decay}
\label{sec:left_constraints}
\begin{figure}
	\centering
	\includegraphics[width=7 cm]{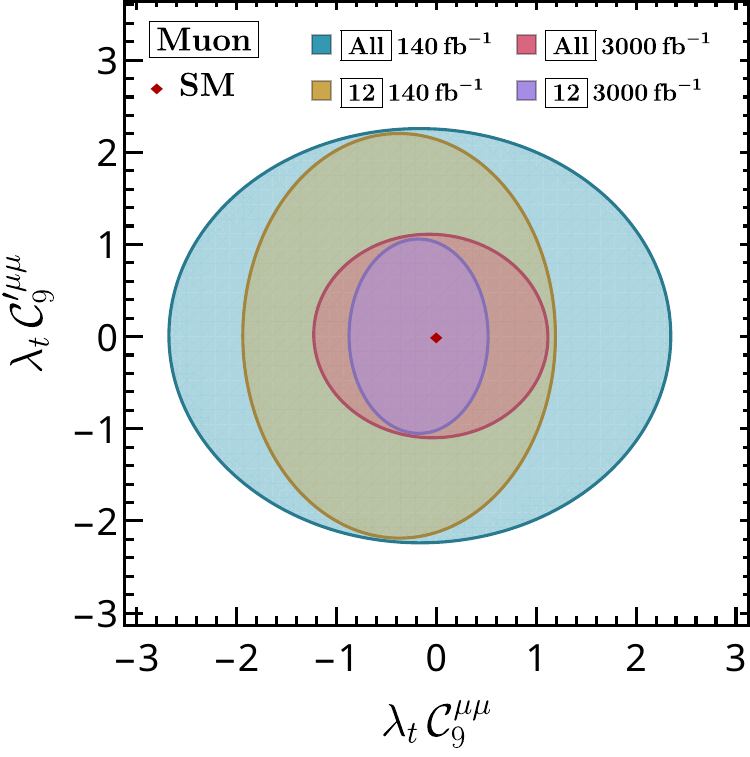}
    \includegraphics[width=7 cm]{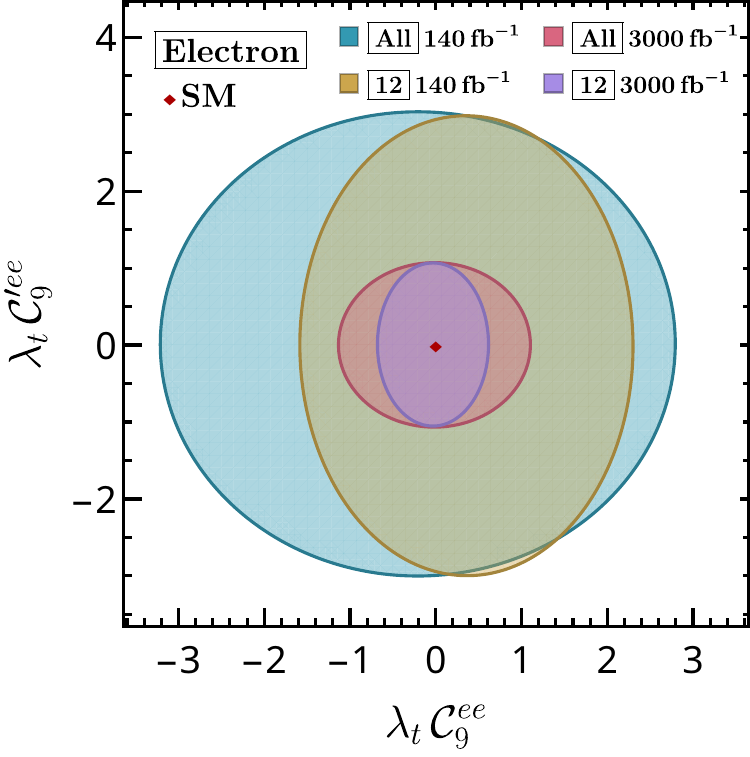}
	\caption{\small Projections of LEFT parameter space mapped from the allowed 95$\%$ CL SMEFT regions for the WCs $\WC_{9}^{(\prime)\mu\mu}$ (left) and $\WC_{9}^{(\prime)ee}$ (right): once with only the flavour indices `12' of the relevant SMEFT WCs setting all others zero (brown, purple), and another including the effects of `all' (`11',`22',`12') of them (blue, green). The two luminosity options are $\mathcal{L}=140~fb^{-1}$ and $3000 ~fb^{-1}$. The label `Muon (Electron)' indicates that only measurements with a muons (electrons) in the final state are used for the processes  $pp\to \ell\ell,\ell\nu$.}
	\label{fig:left1}
\end{figure}

\begin{figure}
	\centering
	\includegraphics[width=7 cm]{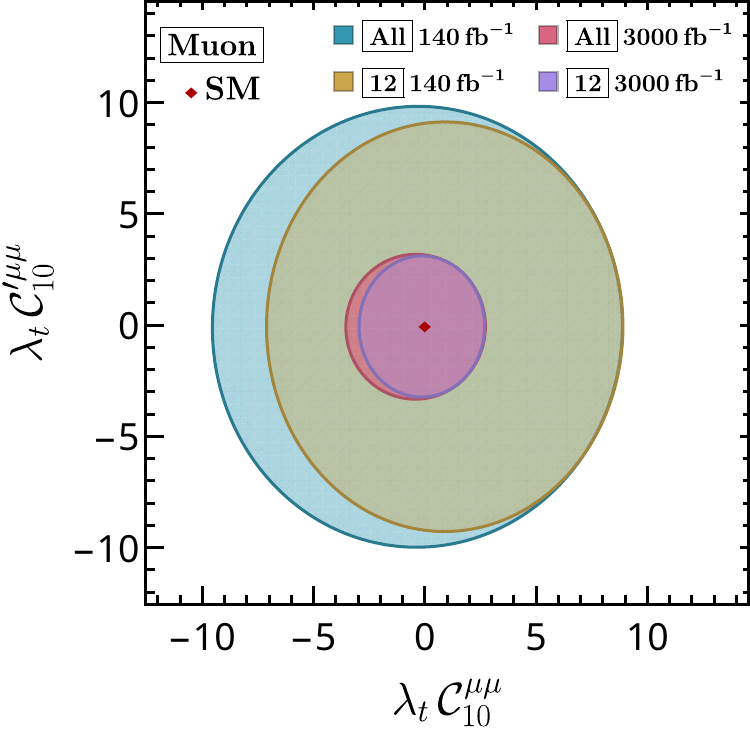}
	\includegraphics[width=7 cm]{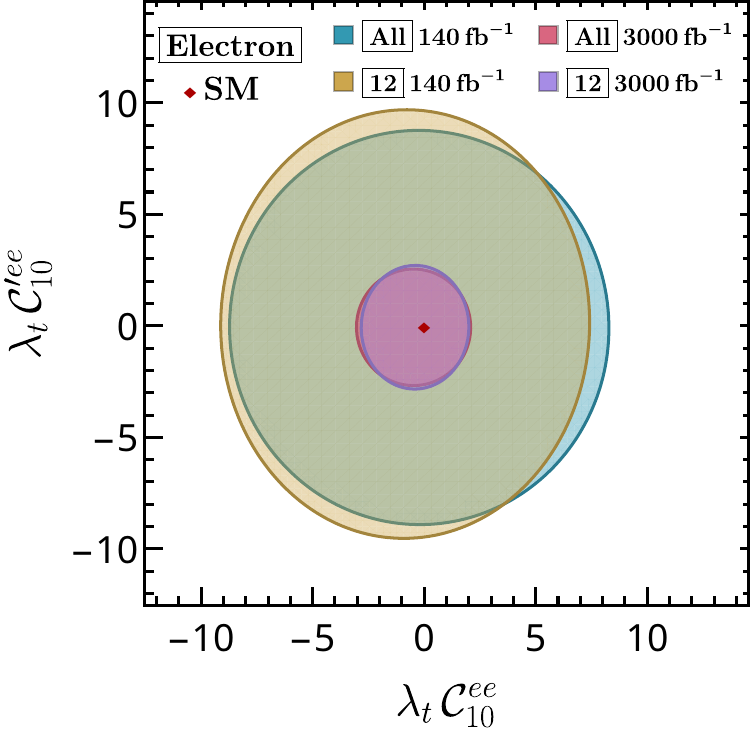}
	\caption{\small  Same as Fig.~\ref{fig:left1} for the WCs $\WC_{10}^{(\prime)\mu\mu}$ (left) and $\WC_{10}^{(\prime)ee}$ (right).}
	\label{fig:left2}
\end{figure}

\begin{figure}
	\centering
	\includegraphics[width=7 cm]{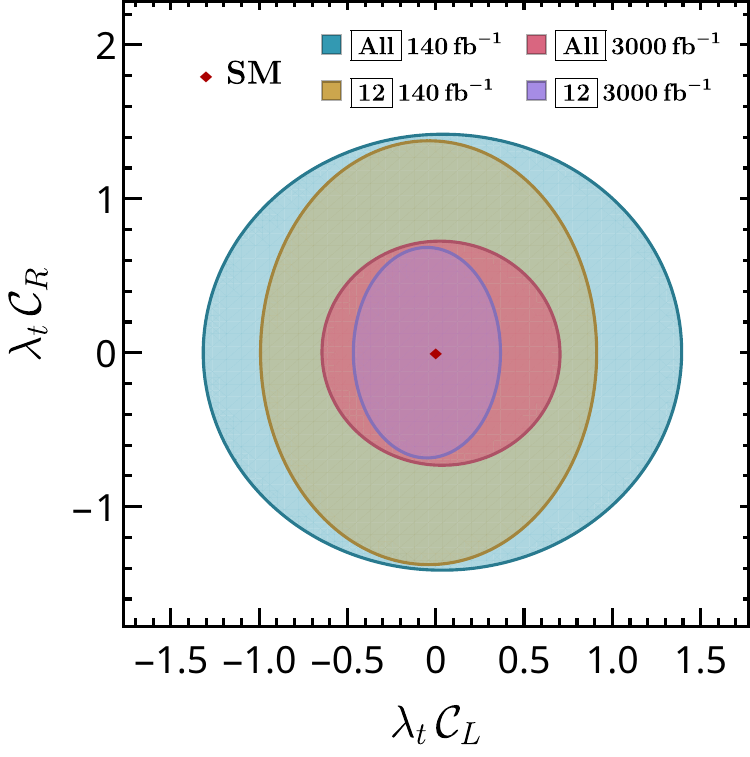}
	\includegraphics[width= 7 cm]{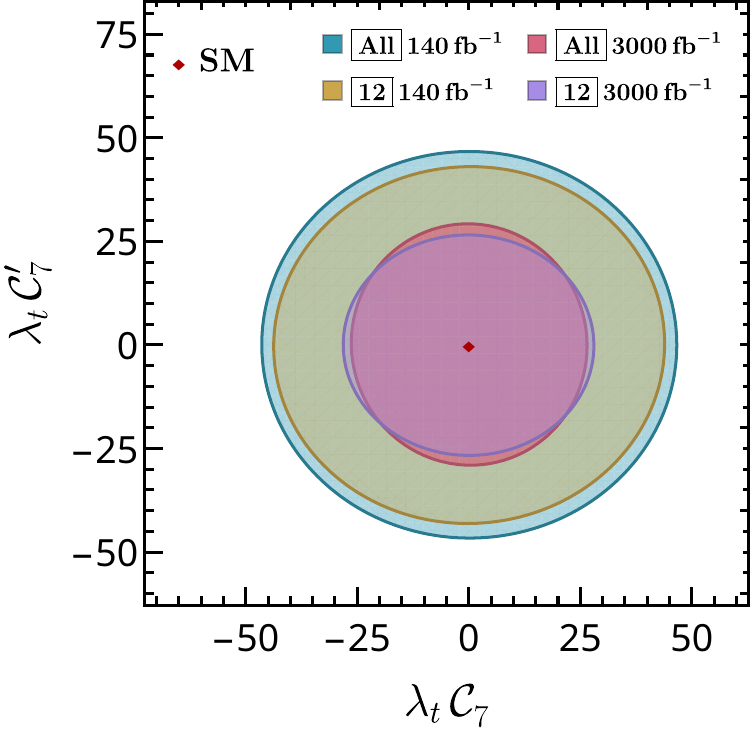}
	\caption{\small Same as Fig.~\ref{fig:left1} for the WCs $\WC_{\rm L,R}$ (left) and $\WC_7^{(\prime)}$ (right).}
	\label{fig:left3}
\end{figure}

Through the matching conditions, Eq.~\ref{eq:matching1}-\ref{eq:matching2}, we can map the allowed parameter regions from the previous section onto regions in the LEFT parameter space,  by reconstructing  the LEFT parameter values for each of the points sampled from the 95\% confidence-level 11~(24)-dimensional SMEFT  region. Note that this reconstruction requires knowledge of the relative weights between operators related by different quark-flavour indices, and that we illustrate only for a couple of possibilities. In Figs.~\ref{fig:left1}-\ref{fig:left3} we show two dimensional projections of the resulting LEFT points for two different scenarios: one in which the UV physics produces only the operators in Eq.~\ref{eq:SMEFT_operators} with the specific $sd$ flavour structure (quark indices `12'); and a second one in with all quark  flavours in the first two generations appear with equal weights (`11', `22', `12'), referred to as $\WC_{i}^{(All)}$.

These two-dimensional projections serve to illustrate several points. First, they show the correlations between the left and right-handed quark currents for vector and axial-vector charged lepton couplings and for left-handed neutrino couplings, respectively.  In addition, operators with muons ($\WC_i^{\mu\mu}$) and those with electrons ($\WC_i^{ee}$) are restricted by different measurements for the processes $pp\to \ell\ell,\ell\nu$. Specifically, we consider final states with muons (electrons) for $\WC_i^{\mu\mu}$ ($\WC_i^{ee}$). Figs.~\ref{fig:left1}-\ref{fig:left2} distinguish between $\WC_i^{\mu\mu}$ (left panels) and $\WC_i^{ee}$ (right panels).

\begin{figure}
	\centering
	\includegraphics[width=7 cm]{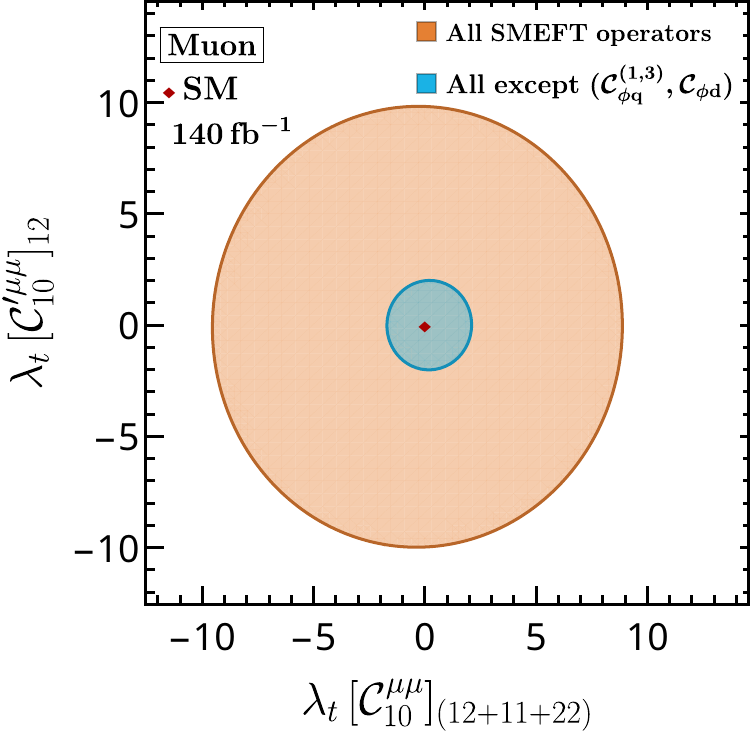}
	\caption{\small  Same as Fig.~\ref{fig:left1}, comparing the contributions of the relevant SMEFT operators to $\WC_{10}^{(\prime)\mu\mu}$: once including all operators (peach) and once with the effects of $\WC_{\phi q}^{(1,3)}, \WC_{\phi d}$ removed (blue) at the current luminosity $\mathcal{L}=140~fb^{-1}$. $\WC_{10}^{\prime}$ does not receive contributions from SMEFT operators with flavor indices ‘11’ and ‘22’ and is therefore presented with the index ‘12’.}
	\label{fig:left_c10_comp}
\end{figure}

The constraints on a LEFT operator are dependent on which SMEFT operators are matched onto them and their respective bounds. For example, while mapping the allowed SMEFT parameter space to LEFT, $\WC_9^{ii}$ is mostly sensitive to the effect of 2Q2L operators, because the effect of the $\rm 2Q^{\prime}2H$ operators is suppressed by a factor $(1-4\sin^2\theta_W)$ (Eq.~\ref{eq:matching1}). 
In contrast, the $\rm 2Q^{\prime}2H$ contributions to $\WC_{10}^{ii}$ are not suppressed (see Eq.~\ref{eq:matching_C10}). In fact, the limits on the $\rm 2Q^{\prime}2H$ SMEFT operators being weaker than those on 2Q2L ones, lead to a weaker bound on $\WC_{10}^{ii}$ when it appears in the matching. Removing the contributions of $\rm 2Q^{\prime}2H$ operators from  $\WC_{10}^{ii}$ results in a stronger bound (and a different meaning of  $\WC_{10}^{ii}$ ), as illustrated in Fig.~\ref{fig:left_c10_comp}. The constraints in the flavour-specific case are slightly stronger wherever the mapping to LEFT is dominated by 2Q2L operators (e.g., $\WC_9, \WC_L$) and comparable in cases where the effects of $\rm 2Q^{\prime}2H$ are significant (e.g., $\WC_{10}$). Additionally, the primed directions ($\WC_i^{\prime}$) are not affected by the flavour-degenerate operators, except for $\WC_7$, and are therefore always comparable in size.

\begin{figure}
    \centering
    \includegraphics[width= 10 cm]{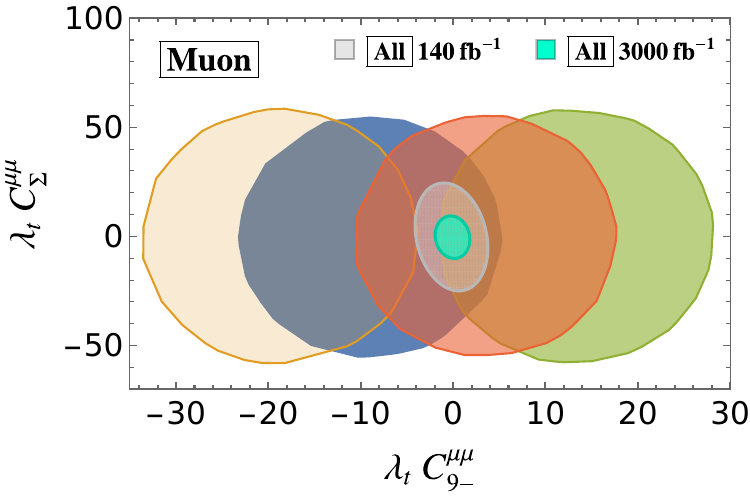}
    \caption{\small LHC constraints on the $\WC_{\Sigma}= (\WC_{10-}-7\WC_{P-})$ vs $\WC_{9-}$ parameter region (grey for $\mathcal{L}=140\;\rm fb^{-1}$ and cyan for $3000\;\rm fb^{-1}$) mapped from the allowed $95\%$ CL 24-dimensional SMEFT region  including the effects of all ($11,22,12$) and the bounds from $\mathcal{B} (\Sigma^+\to p \mu^+\mu-)$ measurements considering four relativistic SM long-distance solutions (blue, yellow, green, red).}
    \label{fig:left6}
\end{figure}

For comparison with the existing bounds obtained from rare kaon and hyperon modes, we also extract the 90\% confidence level intervals for individual Wilson coefficients, which are shown in Table~\ref{tab:onedclLHC} and which can be contrasted to the those on Table~\ref{tab:constraints}. In comparing these two Tables, it is important to keep in mind that they do not quite correspond to the same quantities. For the high-${\rm p_T}$ observables, the interval represents only the experimental uncertainties.

For the low-energy processes, on the other hand, the bounds shown on the third column of Table~\ref{tab:constraints} reflect {\it exclusively} the 90\% experimental result displayed in the second column. In particular, the long-distance contributions to these decay modes are taken at face value from the cited literature.
\begin{table}[]
    \centering
	\begin{tabular}{ |c|c|c|c|c||c|c|c|c|c|}
		\hline
        \multicolumn{5}{|c||}{Muon} & \multicolumn{5}{c|}{Electron}\\ \hline
        Luminosity $\to$& \multicolumn{2}{c}{$140\;fb^{-1}$}  &\multicolumn{2}{|c||}{$3000\;fb^{-1}$}& Luminosity $\to$ &\multicolumn{2}{c}{$140\;fb^{-1}$}  &\multicolumn{2}{|c|}{$3000\;fb^{-1}$}  \\\hline
Parameter $\downarrow$ & Min & Max & Min & Max & Parameter $\downarrow$ & Min & Max & Min & Max \\ \hline
$\lt\, \WC_{9+}^{\mu\mu}$ & -1.7 & 1.3 & -0.8 & 0.7 &$\lt\, \WC_{9+}^{ee}$ & -2.1 & 1.7 & -0.7 & 0.7 \\ \hline
$\lt\, \WC_{9-}^{\mu\mu}$ & -1.7 & 1.3 & -0.9 & 0.7 & $\lt\, \WC_{9-}^{ee}$ & -2.1 & 1.7 &  -0.7 & 0.7  \\ \hline
$\lt\, \WC_{10+}^{\mu\mu}$ & -6.5 & 5.7  & -2.4 & 1.4 & $\lt\, \WC_{10+}^{ee}$ & -5.5 & 4.9  &  -2.0 & 0.9 \\ \hline
$\lt\, \WC_{10-}^{\mu\mu}$ & -6.3 & 5.8  & -2.2 & 1.6 & $\lt\, \WC_{10-}^{ee}$ & -5.4 & 5.1  &  -1.8 & 1.0 \\ \hline
$\lt\, \WC_{7+}^{}$ & -46 & 46 & -25 & 25 & $\lt\, \WC_{7+}^{}$ & -39 & 39 &  -24 & 24  \\ \hline
$\lt\, \WC_{7-}^{}$ & -46 & 46 & -25 & 25 & $\lt\, \WC_{7-}^{}$ & -39 & 40 &  -24 & 24  \\ \hline
$\lt\, \WC_{\nu\nu+}^{}$ & -0.8 & 0.8 & -0.4 & 0.5 & $\lt\, \WC_{\nu\nu+}^{}$ & -1.1 & 1.1 & -0.4& 0.4 \\ \hline
$\lt\, \WC_{\nu\nu-}^{}$ & -0.8 & 0.8 & -0.4 & 0.5 & $\lt\, \WC_{\nu\nu-}^{}$ & -1.1 & 1.1 & -0.4 & 0.4 \\ \hline
$\lt\, (\WC_{S}^{\mu\mu}+\WC_{P}^{\mu\mu})$ & -1.2 & 1.2 & -0.53 & 0.52 & $\lt\, (\WC_{S}^{ee}+\WC_{P}^{ee})$ & -1.5 & 1.5 & -0.57 & 0.57 \\ \hline
$\lt\, (\WC_{S}^{\mu\mu}-\WC_{P}^{\mu\mu})$ & -3 & 3 & -1.2 & 1.2 & $\lt\, (\WC_{S}^{ee}-\WC_{P}^{ee})$ & -2.1 & 2.1 & -1.1 & 1.1 \\ \hline
$\lt\, (\WC_{S}^{\prime\mu\mu}+\WC_{P}^{\prime\mu\mu})$ & -1.7 & 1.7 & -0.7 & 0.7 & $\lt\, (\WC_{S}^{\prime ee}+\WC_{P}^{\prime ee})$ & -1.6 & 1.6 & -0.7 & 0.7 \\ \hline
$\lt\, (\WC_{S}^{\prime\mu\mu}-\WC_{P}^{\prime\mu\mu})$ & -1.7 & 1.6 & -0.66 & 0.66 & $\lt\, (\WC_{S}^{\prime ee}-\WC_{P}^{\prime ee})$ & -1.2 &  1.1 & -0.65 & 0.65 \\ \hline
\end{tabular}
 \caption{\small One-dimensional projections of the allowed 90$\%$ CL ranges for individual LEFT  WC  including the effects of all (`11', `22', `12') the SMEFT operators with different quark-flavours. Bounds on $\WC_{7\pm}$ and $\WC_{\nu\nu\pm}$  are presented separately for the measurements of Muon and Electron final-states at the LHC.}
\label{tab:onedclLHC}
\end{table}

Table~\ref{tab:constraints} also lists bounds on two combinations of $\WC_{10-}$ and $\WC_{P-}$, \,$\WC_{KL}\equiv \WC_{10-}+9.75~\WC_{P-},~\WC_{\Sigma}\equiv (\WC_{10-}-7\WC_{P-})$. These two combinations illustrate the complementarity between kaon and hyperon modes as discussed in \cite{Roy:2024hqg}, with the kaon modes being sensitive only to $\WC_{KL}$. Additional information on $\WC_{\Sigma}$ is shown in Fig.~\ref{fig:left6} as taken from \cite{Roy:2024hqg}. For the kaon modes with neutrinos, the quoted numbers assume neutrino flavour universality. 

Table~\ref{tab:constraints} is split into four groups as follows. The first block refers to kaon modes with charged leptons that constrain only one LEFT coefficient at a time, with results marked $^{\color{red} \star}$ corresponding to lepton universality violation assuming all new physics occurs in the muon modes and were obtained in \cite{Roy:2024hqg}. The second block lists the five coefficients that are obtained from simultaneously considering the observables listed in the first column and the one-parameter limits of the first block \cite{Roy:2024hqg}. After that, we show the constraints from modes with neutrinos, which are theoretically very clean. The last block with the photon dipole coefficients is taken from \cite{Mertens:2011ts}.

\begin{table}[t] \setlength{\tabcolsep}{2ex}
\begin{tabular}{|c|c|c|} \hline
Observable & 90\% CL range & Constraint on WC \\ \hline
${\cal B}(K_L\to \mu^+\mu^-)$ \cite{ParticleDataGroup:2024cfk}& $(6.66,7.02)\times 10^{-9}$ &  $-2.3\times 10^{-3}\leq \lambda_t ~\WC_{KL}^{\mu\mu}\leq-1.7\times 10^{-3}$\\ & & ${\rm ~~or~~}  {-}3.5\times 10^{-4}\leq \lambda_t ~\WC_{KL}^{\mu\mu}\leq 2.6\times 10^{-4}$\\
${\cal B}(K_L\to e^+e^-)$\cite{ParticleDataGroup:2024cfk} & $(2,18)\times 10^{-12}$ &
$-7\times 10^{-3}\leq \WC_{P-}^{ee}\leq 1.6\times 10^{-3}$\\
${\cal B}(K_S\to \mu^+\mu^-)$\cite{LHCb:2020ycd}& $<2.1\times 10^{-10} $& $|\lambda_t ~\WC_{S-}^{\mu\mu}|<1.7\times 10^{-3}$\\
${\cal B}(K_L\to \pi^0\mu^+\mu^-)$\cite{AlaviHarati:2000hs}& $< 3.8\times 10^{-10}$ & $|\lambda_t ~\WC_{S+}^{\mu\mu}| < 6.4\times 10^{-4} $\\
${\cal B}(K_L\to \pi^0e^+e^-)$\cite{AlaviHarati:2003mr}& $< 2.8\times 10^{-10}$ & $|\lambda_t ~\WC_{S+}^{ee}| < 2.6\times 10^{-4}$\\
$a_S^{\mu\mu}-a_S^{ee}$\cite{Batley:2003mu,Batley:2004wg} & $(-0.3,1.3)$ &  $-0.21<\lambda_t \WC_{9+}^{\mu\mu}<0.92 ^{\color{red} \star}$\\
$ |a_+^{\mu\mu}-a_+^{ee}|$ \cite{NA482:2009pfe,NA62:2022qes} & $< 0.035$ &  $|\lambda_t\WC_{9+}^{\mu\mu}| < 2.5\times 10^{-2}\vphantom{|_|^|} ^{\color{red} \star}$\\ \hline
from kaon-hyperon  fit\cite{Roy:2024hqg}&  & solution 4 \\ \hline
$10^8~\Delta{\cal B}(K^+\to\pi^+\mu^+\mu^-)^{\color{red} \star}$ & $(-1.2,0.4)$ & $|\lambda_t ~\WC_{10+}^{\mu\mu}| < 0.63$ \\
$\big|A_{{\rm FB},K^+}(K^+\to\pi^+\mu^+\mu^-)|$ & $<0.12$ & $|\lambda_t ~\WC_{10-}^{\mu\mu}| < 32$ \\
$10^8~{\cal B}(\Sigma^+ \to p \mu^+\mu^-)$\cite{LHCb:2017rdd,LHCb:2024fhb} & $(0.26,5.2)$ &$|\lambda_t ~\WC_{P+}^{\mu\mu}| < 0.2$\\
&&$|\lambda_t ~\WC_{P-}^{\mu\mu}| < 0.4$\\
&& $-10.6\leq \lambda_t ~\WC_{9-}^{\mu\mu} < 17.7$ \\
&& $|\lambda_t ~\WC_{\Sigma}^{\mu\mu}|< 55 $ \\ \hline
modes with neutrinos &&\\ \hline
${\cal B}(K^+\to \pi^+\nu\bar\nu)$ \cite{NA62:2020fhy,NA62:2021zjw}&   $(4.8,17.3)\times 10^{-11}$ & $-2.9\times 10^{-3}\leq \lambda_t ~\WC_{\nu\nu+}\leq 5.1\times 10^{-3}$ \cite{He:2018uey} \\
${\cal B}(K_L\to \pi^0\pi^0\nu\bar\nu)$ \cite{E391a:2011aa}& $<8.1\times10^{-7}$ & $-5.4\leq \lambda_t ~\WC_{\nu\nu-}\leq 5.4 $  \cite{Geng:2021fog}\\
\hline
fit from \cite{Mertens:2011ts}  & Result from \cite{Mertens:2011ts} & Constraint on WC\\ \hline
$K\to \pi \pi\gamma$ \cite{Mertens:2011ts} & $ | C^-_\gamma|\lesssim 0.1 G_{\rm F}m_K$ &  $|\lambda_t~C_{7-}|<9.8\times 10^{-2}\vphantom{|_|^{|^|}}$ \\
$K\to \gamma\gamma$ \cite{Mertens:2011ts}& $| C^{\pm}_\gamma|\lesssim 0.3 G_{\rm F}m_K$  &  $|\lambda_t~C_{7\pm}|<0.29\vphantom{|_|^|}$ \\ \hline
\end{tabular}
\caption{Constraints on one WC at a time from kaon and hyperon observables. All experimental measurements and limits are taken at 90\%-CL but the theory uncertainty is not quantified as discussed in the text. We first list modes that constrain only one coefficient for charged leptons, followed by constraints from the overall fit of \cite{Roy:2024hqg}, modes with neutrinos, and finally, results from the fit of \cite{Mertens:2011ts}. Results marked $^{\color{red} \star}$ correspond to lepton universality violation assuming all new physics occurs in the muon modes. The two combinations of coefficients  listed are 
\,$\WC_{KL}\equiv \WC_{10-}+9.75~\WC_{P-},~\WC_{\Sigma}\equiv (\WC_{10-}-7\WC_{P-})$.}
\label{tab:constraints}
\end{table}

Looking at future prospects, Table~\ref{tab:onedclLHC} includes both the current limits from LHC as well as projections for 3000~fb$^{-1}$ assuming rescaled background uncertainty (see Sec.~\ref{sec:uncertainties}). The situation for kaon and hyperon modes into charged leptons is less clear, as the dominant uncertainty lies in long-distance contributions. For illustration, we consider a scenario in which these are known to within 10\% for $K^+\to\pi^+\mu^+\mu^-$ from lattice studies \cite{Anzivino:2023bhp} and assume something similar can eventually be achieved for $\Sigma^+\to p\mu^+\mu^-$ and that in both cases, the experimental uncertainty can reach this level or better (it already has for the kaon mode). Using solution  IV of \cite{Roy:2024hqg} for the hyperon mode, we present two possible figures of merit in Table~\ref{tab:futkaons} taking one non-zero WC at a time in both cases. The first two columns are the limit that the kaon and hyperon mode, respectively, could place under the stated assumptions. The last two columns display 
the relative change in the two branching ratios induced by NP with the Wilson coefficient taking its projected upper bound from LHC with 3000~fb$^{-1}$.
 \begin{table}[]
    \centering
	\begin{tabular}{|c|c|c|c|}
\hline
$\frac{\Delta{\cal B}(\WC_i)}{{\cal B}_{\rm exp}}(K^+\to \pi^+\mu^+\mu^-)\leq 10\% $ & $\frac{\Delta{\cal B}(\WC_i)}{{\cal B}_{\rm exp}}(\Sigma^+\to p \mu^+\mu^-)\leq 10\% $ 
& $\frac{\Delta{\cal B}_{K^+}(\WC_{i-{\rm max}})}{{\cal B}_{\rm exp}}$ &$\frac{\Delta{\cal B}_{\Sigma^+}(\WC_{i-{\rm max}})}{{\cal B}_{\rm exp}}$ \\ \hline
$|\lt\, \WC_{9+}^{\mu\mu}|\lesssim 0.03$ & $|\lt\, \WC_{9+}^{\mu\mu}|\lesssim 4$ &5& 0.009\\ 
 -& $|\lt\, \WC_{9-}^{\mu\mu}|\lesssim 7$ &-& 0.1\\
$|\lt\, \WC_{10+}^{\mu\mu}|\lesssim 0.1$ & $|\lt\, \WC_{10+}^{\mu\mu}|\lesssim 1.5$ &32& 0.2\\
-& $|\lt\, \WC_{10-}^{\mu\mu}|\lesssim 5$ &-& 0.014\\
$|\lt\, \WC_{7+}^{}|\lesssim 0.1$ & $|\lt\, \WC_{7+}^{}|\lesssim 2$&200& 2.5\\
-& $|\lt\, \WC_{7-}^{}|\lesssim 8$&-& 6\\
$|\lt\, \WC_{S+}^{\mu\mu}|\lesssim 4$ &$|\lt\, \WC_{S+}^{\mu\mu}|\lesssim 14$ &0.002 & $1\times 10^{-4}$\\
-& $|\lt\, \WC_{S-}^{\prime\mu\mu}|\lesssim 96$ &-&$1\times 10^{-5}$\\
$|\lt\, \WC_{P+}^{\prime\mu\mu}|\lesssim 3$ & $|\lt\, \WC_{P+}^{\prime\mu\mu}|\lesssim 6$ &0.007& 0.001\\
-& $|\lt\, \WC_{P-}^{\prime\mu\mu}|\lesssim 35$  &-&$4\times 10^{-5}$\\ \hline
\end{tabular}
\caption{Illustrative future projections of the limits that could be obtained from ${\cal B}(K^+\to \pi^+\mu^+\mu^-)$ and ${\cal B}(\Sigma^+\to p \mu^+\mu^-)$. The two left columns assume a $10\%$ uncertainty on the long-distance calculations. The two rightmost columns present the relative changes in ${\cal B}$ resulting from the corresponding WC taking its upper  limit from the HL-LHC projections in Table~\ref{tab:onedclLHC}.}
\label{tab:futkaons}
\end{table}

The numbers in Tables~\ref{tab:onedclLHC}-\ref{tab:futkaons} display the complementarity
between the different experiments. Salient points are: 
\begin{itemize}
    \item The kaon bounds, when they exist, are already more stringent than those from hyperons or high-${\rm p_T}$ observables. The caveat is that, except for the modes with neutrinos, the uncertainty in long-distance contributions will not be reliably estimated until lattice results become available.
    \item  The kaon modes measured thus far do not constrain the combination $\WC_\Sigma$ nor the direction $\WC_{9-}$ where the hyperons are complementary but provide much weaker bounds than high-${\rm p_T}$ observables. For electron modes, lepton mass suppression implies that $K_L\to e^+ e^-$ is mostly  sensitive to $\WC_{P-}^{ee}$ and not to $\WC_{10-}^{ee}$ in the combination $\WC_{KL}$. 
    \item The relative weakness of the kaon/hyperon bounds for $\WC_{9-,10-}$ indicates, for example, that high-${\rm p_T}$ processes are presently better at constraining models with new $Z^\prime$ gauge bosons with only axial couplings \cite{Ismail:2016tod}.
    \item  For scalar operators, the kaon constraints are dominant for $\WC_{S\pm}^{\mu\mu}$ and $\WC_{S+}^{ee}$. The high-${\rm p_T}$ observables are more important for  $\WC_{S-}^{ee}$ and $\WC_{P\pm}$. However, the distinction between $\WC^{(\prime)}_{S,P}$ appearing in the $\WC_{S,P,\pm}$ combinations occurs only at dimension eight, where there are many other possible SMEFT operators not discussed in this study.
    \item Prospects for new kaon measurements are currently limited to NA62 and KOTO with the neutrino modes. NA62 results already  place the most stringent constraints on $\WC_{\nu\nu+}$, and their sensitivity from the full data set is  expected to improve this by at most a factor of two. Its counterpart,  $\WC_{\nu\nu-}$, is much poorly constrained at present from $K_L\to \pi^0\pi^0\nu\bar\nu$  \cite{E391a:2011aa}. To compete with the expected constraint from high-${\rm p_T}$ observables with 3000 fb$^{-1}$,  $|\WC_{\nu\nu-}|\lesssim 1$, the upper limit on the branching ratio  ${\cal B}(K_L\to \pi^0\pi^0\nu\bar\nu)$ would need to improve by a factor of about 600, implying a single event sensitivity of $\sim 5 \times 10^{-10}$ which is five times better than what is expected from KOTO \cite{Goudzovski:2022vbt}. 
    \item More generically, the constraints from high-${\rm p_T}$ observables depend on many more parameters than each kaon mode, limiting their significance to specific models or simple scenarios where only a few SMEFT coefficients are involved. The low-energy observables (except for the neutrino modes), on the other hand, suffer from systematic uncertainties due to long-distance effects with no immediate prospects of being resolved by lattice QCD.
    \item All this, plus the very different experimental issues affecting these two types of observables, guarantees that the cross-talk between the two approaches will play an important role in constraining the type of new physics we discuss. 
\end{itemize}

\section{Discussion and comparison with the literature}
\label{sec:comparison}

\subsection{Importance of including full set of SMEFT operators}

\begin{figure}
    \centering
    \includegraphics[width=7 cm]{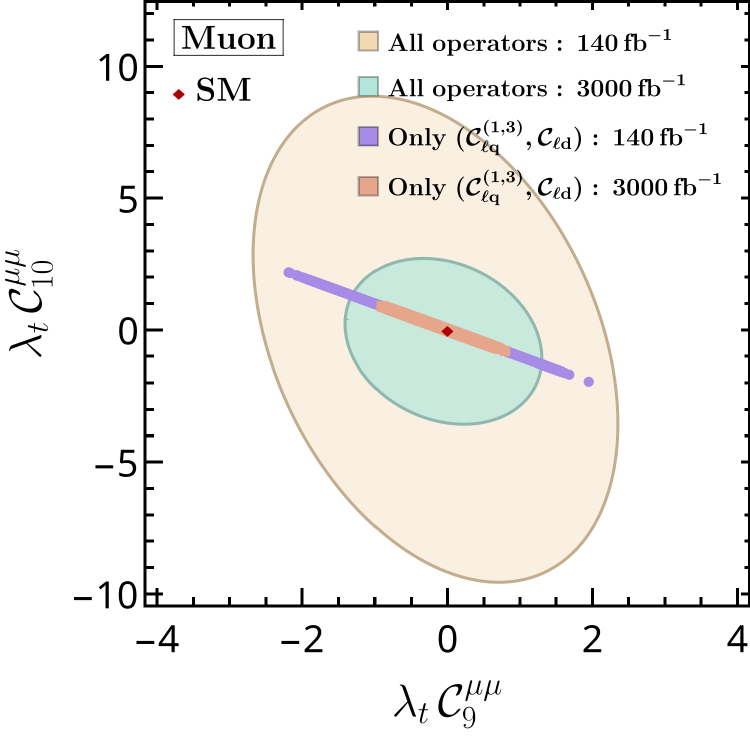}
    \caption{\small Projections of the allowed 95$\%$ CL regions at $\mathcal{L}=140, 3000\;\rm fb^{-1}$ using full set of SMEFT operators (peach, aqua) and using only a sub-set --- $\oplqone,\oplqthree,\opld$ ( violet, coral) for $\WC_9^{\mu\mu},\WC_{10}^{\mu\mu}$. }
    \label{fig:left_c9c10}
\end{figure}

Previous studies ~\cite{Fajfer:2023nmz,Hiller:2024vtr} based on the four fermion operators have neglected 2Q2H-type SMEFT operators as well as operators involving right-handed lepton singlets when matching to LEFT.  By including them in this study, we have shown that  these operators play a significant role in constraining vector-vector or vector-axialvector low-energy operators $\mathcal{O}_9$, $\mathcal{O}_9^{\prime}$, $\mathcal{O}_{10}$, $\mathcal{O}_{10}^{\prime}$.  Fig.~\ref{fig:left_c10_comp} and Fig.~\ref{fig:left_c9c10} illustrate the impact of the operators $\opphiqone$, $\opphiqthree$, $\opphid$, and $\opqe$ on $\WC_9$, $\WC_9$, and $\WC_{10}^{\prime}$. Without these additional degrees of freedom, $\WC_9$ and $\WC_{10}$ become dependent (with $\WC_9 = -\WC_{10}$), as shown in Fig.~\ref{fig:left_c9c10}, and the same applies to $\WC_9^{\prime}$ and $\WC_{10}^{\prime}$. Fig.~\ref{fig:left_c10_comp} highlights the case for $\WC_{10}^{(\prime)}$, where the distinct effects of $\wcphiqone + \wcphiqthree$ (associated with $\WC_{10}$) and $\wcphid$ (associated with $\WC_{10}^{\prime}$) operators are evident.

\subsection{Importance of VH measurements}

\begin{figure}
    \centering
    \includegraphics[width=7 cm]{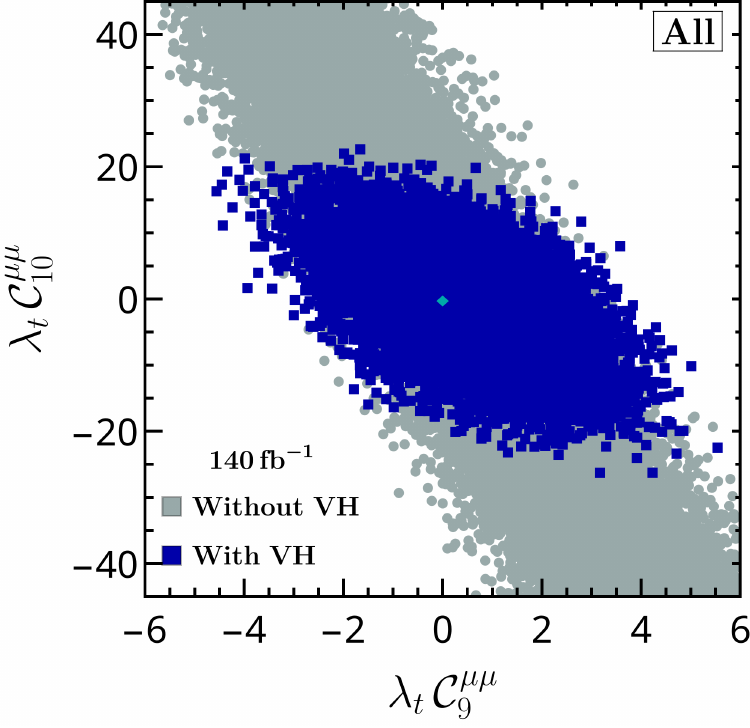}
    \includegraphics[width= 7 cm]{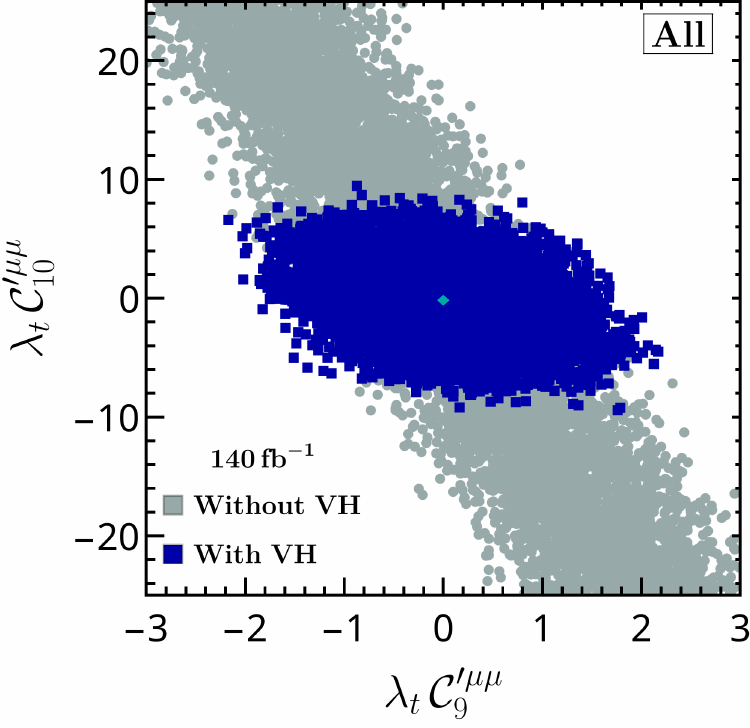}
    \caption{\small Projections of the allowed 95$\%$ CL regions at $\mathcal{L}=140\;\rm fb^{-1}$ considering the effects of VH measurements (blue) or excluding them (grey) for the LEFT WCs $\WC_9^{\mu\mu},\WC_{10}^{\mu\mu}$(left),  $\WC_9^{\prime\mu\mu},\WC_{10}^{\prime\mu\mu}$ (right). }
    \label{fig:left_VH}
\end{figure}

The inclusion of VH measurements is essential for constraining the 2Q2H-type operators, which in turn affect the low-energy operators $\WC_9^{(\prime)}$ and $\WC_{10}^{(\prime)}$. Fig.~\ref{fig:left_VH} illustrates the impact of VH measurements on $\WC_9^{(\prime)}$ and $\WC_{10}^{(\prime)}$.

\subsection{Scalar operators}\label{sec:scalar}

\begin{figure}
    \centering
    
    \includegraphics[width= 7 cm]{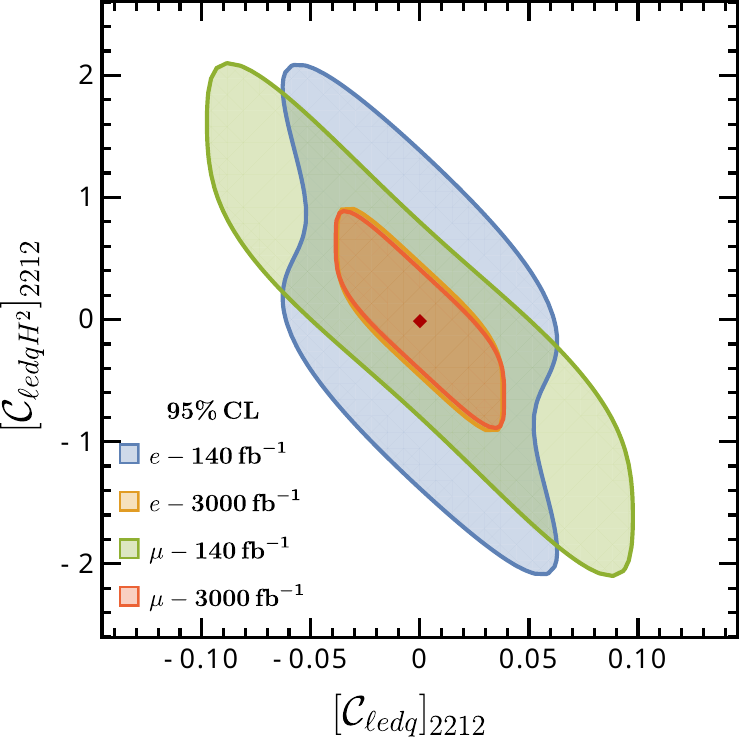}
    \caption{\small Same as Fig.~\ref{fig:smeft_95CL1} for the SMEFT WCs $[\wcledq]_{2212}$ (dimension-6) and $[\WC_{\ell e q d H^2}^{(3)}]_{2212}$ (dimension-8).}
    \label{fig:smeft_95CL5}
\end{figure}

\begin{figure}
    \centering
    \includegraphics[width= 7 cm]{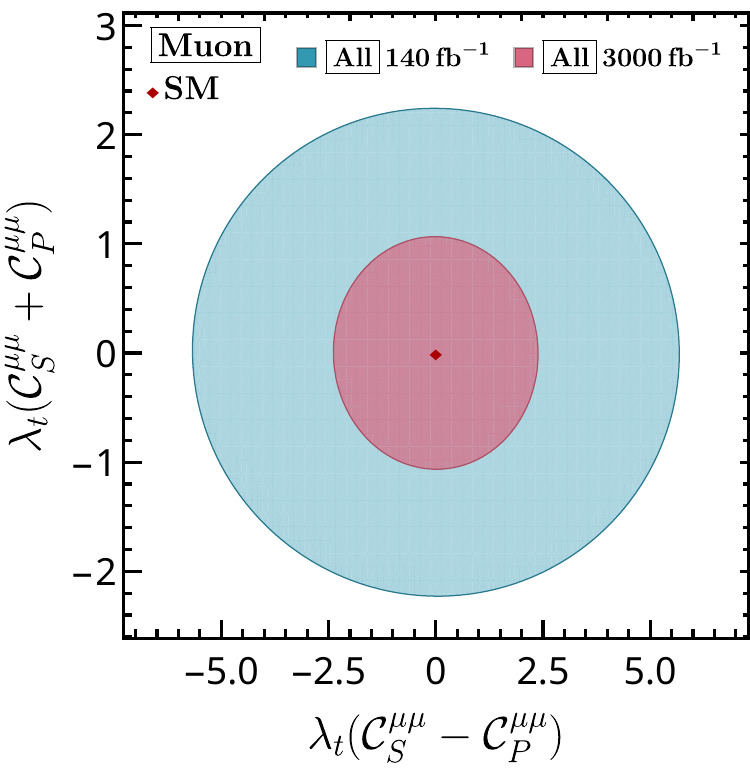}
    \includegraphics[width= 7 cm]{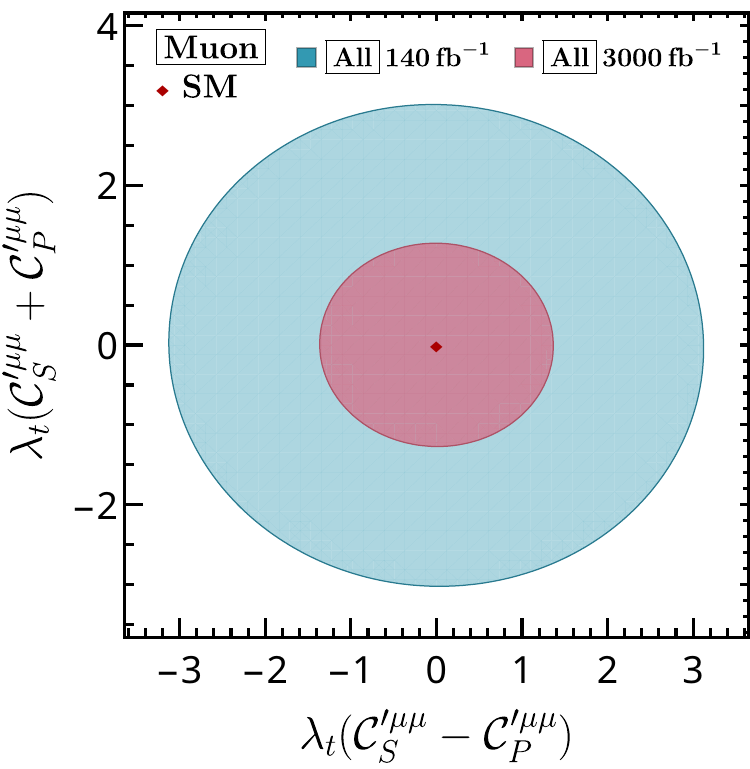}
    \caption{\small Same as Fig.~\ref{fig:left1} for the WC combinations $\WC_S^{\mu\mu}\pm \WC_P^{\mu\mu}$ (left) and $\WC_S^{\prime\mu\mu}\pm \WC_P^{\prime\mu\mu}$ (right) including the effects of all SMEFT operators with indices `11', `12', `22'. The directions $\WC_S^{\mu\mu}+ \WC_P^{\mu\mu}$ and $\WC_S^{\prime\mu\mu}- \WC_{P}^{\prime\mu\mu}$ singles out the dimension-8 contributions, which are otherwise unbounded.}
    \label{fig:left7}
\end{figure}

As mentioned earlier, the contributions of dim-6 SMEFT operators do not differentiate between $\WC^{(\prime)}_S$ and $\WC^{(\prime)}_P$, but this distinction does arise at dim-8. To illustrate what happens in the case of dilepton and monolepton searches, we can use the example of Eq.~\ref{dim8m} along with $\WC_{\ell edq}$. Expanding the two operators, we see that:
\begin{align}
	\Op_{\ell e d q}[2212], ~\Op_{\ell e d q}[2212]^\dagger \to & \left((\bar \nu P_R \mu)(\bar d P_L c)+(\bar\mu P_R \mu) (\bar d P_L s) \right),\\ 
&	\left((\bar \mu P_L \nu)(\bar c P_R d)+(\bar\mu P_L \mu) (\bar s P_R d) \right) \\
  \Op_{\ell e d q}[2221] ,~\Op_{\ell e d q}[2221]^\dagger\to & \left((\bar \nu P_R \mu)(\bar s P_L u)+(\bar\mu P_R \mu) (\bar s P_L d) \right),\\
&	 \left((\bar \mu P_L \nu)(\bar u P_R s)+(\bar\mu P_L \mu) (\bar d P_R s) \right) \\
	\Op^{(3)}_{\ell e d q H^2}[2212],~\Op^{(3)}_{\ell e d q H^2}[2212]^\dagger \to &(\bar\mu P_R \mu) (\bar d P_R s) 
	,~ (\bar\mu P_L \mu) (\bar s P_L d)  \\
 \Op^{(3)}_{\ell e d q H^2}[2221],~ \Op^{(3)}_{\ell e d q H^2}[2221]^\dagger \to & (\bar\mu P_R \mu) (\bar s P_R d),~
	 (\bar\mu P_L \mu) (\bar d P_L s),
\end{align}
 where we have considered only the first and second generations of quarks and the second generation of leptons for simplicity. This implies that the two can be separated by considering the monolepton and dilepton results separately. The resulting bounds are shown in Fig.~\ref{fig:smeft_95CL5}. The constraint on $\WC_S^{(\prime)}\pm\WC_P^{(\prime)}$ arises from dim-8 and dim-6, respectively, and the other way around for the primed coefficients; the results are displayed in Fig.~\ref{fig:left7}.

\section{Summary and outlook}
\label{sec:summary}

We have considered the high-$\rm p_T$ production of dilepton, monolepton, monojet and VH at LHC in order to constrain the dimension-six SMEFT operators that affect rare kaon and hyperon decay. 
The dilepton and monolepton searches place the strongest constraints on the four-fermion operators and are the only ones sensitive to $\Op_{\ell edq}$, $\Op_{ed}$ and $\Op_{qe}$. 
The $\Op_{2Q^{\prime}2H}$ operators receive their strongest constraints from the VH searches.  $\mathcal{O}_{d\gamma}$ is constrained by dilepton searches but not VH or monojets and is the direction with the weakest constraints. 

The high-$\rm p_T$ observables provide comparable constraints on the LEFT operators for both muon and electron modes. We find that these constraints are generally weaker than existing ones from rare kaon decay, but they complement them by restricting the directions in parameter space to which kaons are blind. They are also complementary in the sense that they depend on very different sets of assumptions. Constraints extracted from high-${\rm p_T}$ observables apply only to UV complete models that match onto a limited subset of dimension six effective operators. We have illustrated the effect of other operators contributing to the same observables by expanding the quark flavours in our operator subset. Other theoretical uncertainties in this case are quantifiable and include higher order effects  (higher dimension operators or loop contributions) and MC sampling errors. On the other hand, the largest uncertainty with the low-energy observables with charged leptons (and photons) arises from the large long-distance contributions. At present, these are based on hadronic models where uncertainty estimates are unreliable, and progress awaits lattice calculations.

\vspace{2 cm}
\acknowledgments
This work is supported by an Australian Research Council Discovery Project.

\appendix

\section{Experimental results and fitted cross-sections for VH processes}\label{appvh}

For our VH analysis we use the results from \cite{ATLAS:2020jwz} reproduced in Table~\ref{tab:ATLAS_VH}.
\begin{table}
\begin{center}
\begin{tabular}{l|r|cc}
\toprule
STXS region $(|y_H| < 2.5, H \rightarrow b\bar{b})$ & SM prediction (fb)  & Measurement (fb) & (Tot. Unc.)  \\
\toprule
(1) $W \rightarrow \ell \nu;~ p_\textrm{T}^{W,t} \in [250, 400]\,$GeV & 5.83 $\pm$ 0.26~~~~ & 3.3 &$^{+4.8}_{-4.6}$  \\  \hline
(2) $W \rightarrow \ell \nu;~ p_\textrm{T}^{W,t} \in [400, \infty]\,$GeV & 1.25 $\pm$ 0.06~~~~ & 2.1 &$^{+1.2}_{-1.1}$ \\ \hline
(3) $Z \rightarrow \ell \ell, \nu\nu;~ p_\textrm{T}^{Z,t} \in [250, 400]\,$GeV & 4.12 $\pm$ 0.45~~~~ & 1.4 &$^{+3.1}_{-2.9}$  \\ \hline
(4) $Z \rightarrow \ell \ell, \nu\nu;~ p_\textrm{T}^{Z,t} \in [400, \infty]\,$GeV & 0.72 $\pm$ 0.05~~~~ & 0.2 &$^{+0.7}_{-0.6}$  \\
\bottomrule
\end{tabular}
\end{center}
\caption{Measurements for WH and ZH at the ATLAS experiment at $\sqrt{s}=13~\rm TeV$ and luminosity $\mathcal{L}=139~\rm fb^{-1}$\;\cite{ATLAS:2020jwz}.}
\label{tab:ATLAS_VH}
\end{table}

Our fits to the cross-section attributable to NP at LO, $\sigma-\sigma_{SM}$, as calculated with {\tt Madgraph} for the four regions in Table~\ref{tab:ATLAS_VH} are
\begin{align}
\sigma_{WH}(1) \,[fb]&=-6.4[\WC_{\phi q}^{(3)}]_{21}+61.90[\WC_{\phi q}^{(3)}]_{21}^2 
+ 42.3 [\WC_{\phi q}^{(3)}]_{11} + 84.0 [\WC_{\phi q}^{(3)}]_{11}^2
+ 2.8 [\WC_{\phi q}^{(3)}]_{22} + 5.8 [\WC_{\phi q}^{(3)}]_{22}^2\nonumber\\
&-0.075[\WC_{dW}]_{21}+17.56[\WC_{dW}]_{21}^2 +0.13[\WC_{dW}]_{11}+82.1[\WC_{dW}]_{11}^2 + 0.1[\WC_{dW}]_{22} + 4.7[\WC_{dW}]_{22}^2\nonumber\\
&-20.7[\WC_{\phi q}^{(3)}]_{11}[\WC_{\phi q}^{(3)}]_{12} -0.07 [\WC_{\phi q}^{(3)}]_{11}[\WC_{\phi q}^{(3)}]_{22} -6.8[\WC_{\phi q}^{(3)}]_{12}[\WC_{\phi q}^{(3)}]_{22} \nonumber\\
&-0.18[\WC_{dW}]_{11}[\WC_{dW}]_{12} +0.07 [\WC_{dW}]_{11}[\WC_{dW}]_{22} +0.08[\WC_{dW}]_{12}[\WC_{dW}]_{22}
\end{align}
\begin{align}
\sigma_{WH}(2)\,[fb]&=-3.2[\WC_{\phi q}^{(3)}]_{21}+106.46[\WC_{\phi q}^{(3)}]_{21}^2+ 16.2 [\WC_{\phi q}^{(3)}]_{11} + 145.9 [\WC_{\phi q}^{(3)}]_{11}^2
+ 0.7 [\WC_{\phi q}^{(3)}]_{22} + 5.4 [\WC_{\phi q}^{(3)}]_{22}^2\nonumber\\
&+0.065[\WC_{dW}]_{21}+24.03[\WC_{dW}]_{21}^2+0.1[\WC_{dW}]_{11}+148.3[\WC_{dW}]_{11}^2 + 0.01[\WC_{dW}]_{22} + 4.7[\WC_{dW}]_{22}^2\nonumber\\
&-37.8[\WC_{\phi q}^{(3)}]_{11}[\WC_{\phi q}^{(3)}]_{12} -0.023 [\WC_{\phi q}^{(3)}]_{11}[\WC_{\phi q}^{(3)}]_{22} -9.6[\WC_{\phi q}^{(3)}]_{12}[\WC_{\phi q}^{(3)}]_{22} \nonumber\\
&-0.45[\WC_{dW}]_{11}[\WC_{dW}]_{12} +0.11 [\WC_{dW}]_{11}[\WC_{dW}]_{22} +0.023[\WC_{dW}]_{12}[\WC_{dW}]_{22} 
\end{align}
\begin{align}
\sigma_{ZH}(3)\,[fb]&=-4.3[\WC_{\phi q}^{(1)}]_{21} +38.1[\WC_{\phi q}^{(1)}]_{21}^2-4.2[\WC_{\phi q}^{(3)}]_{21}+38.2[\WC_{\phi q}^{(3)}]_{21}^2 \nonumber\\
&+24.4[\WC_{\phi q}^{(1)}]_{11} +51.8[\WC_{\phi q}^{(1)}]_{11}^2 +24.9[\WC_{\phi q}^{(3)}]_{11}+51.6[\WC_{\phi q}^{(3)}]_{11}^2 \nonumber\\
&+1.0[\WC_{\phi q}^{(1)}]_{22} +3.8[\WC_{\phi q}^{(1)}]_{22}^2 +1.9[\WC_{\phi q}^{(3)}]_{22}+3.8[\WC_{\phi q}^{(3)}]_{22}^2 \nonumber\\
&-0.002[\WC_{\phi d}]_{21}+ 15.5 [\WC_{\phi d}]_{21}^2 \nonumber\\
&-10.7[\WC_{\phi q}^{(1)}]_{11}[\WC_{\phi q}^{(1)}]_{12} +0.65[\WC_{\phi q}^{(1)}]_{11}[\WC_{\phi q}^{(1)}]_{22} -5.3[\WC_{\phi q}^{(1)}]_{12}[\WC_{\phi q}^{(1)}]_{22} \nonumber\\
&-11.3[\WC_{\phi q}^{(3)}]_{11}[\WC_{\phi q}^{(3)}]_{12} +0.6[\WC_{\phi q}^{(3)}]_{11}[\WC_{\phi q}^{(3)}]_{22}  -5.5[\WC_{\phi q}^{(3)}]_{12}[\WC_{\phi q}^{(3)}]_{22} \nonumber\\
&-26.0[\WC_{\phi q}^{(1)}]_{11}[\WC_{\phi q}^{(3)}]_{11} -11.0[\WC_{\phi q}^{(1)}]_{11}[\WC_{\phi q}^{(3)}]_{12} +0.6[\WC_{\phi q}^{(1)}]_{11}[\WC_{\phi q}^{(3)}]_{22} \nonumber\\
&-11.3[\WC_{\phi q}^{(1)}]_{12}[\WC_{\phi q}^{(3)}]_{11}  -8.7[\WC_{\phi q}^{(1)}]_{12}[\WC_{\phi q}^{(3)}]_{12}  -5.5[\WC_{\phi q}^{(1)}]_{12}[\WC_{\phi q}^{(3)}]_{22} \nonumber\\
&+0.8[\WC_{\phi q}^{(1)}]_{22}[\WC_{\phi q}^{(3)}]_{11}  -5.5[\WC_{\phi q}^{(1)}]_{22}[\WC_{\phi q}^{(3)}]_{12}  +2.5[\WC_{\phi q}^{(1)}]_{22}[\WC_{\phi q}^{(3)}]_{22}	\nonumber\\
&-0.008[\WC_{\phi q}^{(1)}]_{21}[\WC_{\phi d}]_{21}-0.026[\WC_{\phi q}^{(3)}]_{21}[\WC_{\phi d}]_{21} \nonumber\\
&-0.005[\WC_{\phi q}^{(1)}]_{11}[\WC_{\phi d}]_{21}-0.14[\WC_{\phi q}^{(3)}]_{11}[\WC_{\phi d}]_{21} \nonumber\\
&-0.001[\WC_{\phi q}^{(1)}]_{22}[\WC_{\phi d}]_{21}-0.012[\WC_{\phi q}^{(3)}]_{22}[\WC_{\phi d}]_{21} \nonumber\\
&+10.77[\WC_{dW}]_{21}^2+3.20[\WC_{dB}]_{21}^2+11.68[\WC_{dW}]_{21}[\WC_{dB}]_{21}
\end{align}
\begin{align}
\sigma_{ZH}(4)\,[fb]&=-1.96[\WC_{\phi q}^{(1)}]_{21}+54.81[\WC_{\phi q}^{(1)}]_{21}^2-1.54[\WC_{\phi q}^{(3)}]_{21}+54.64[\WC_{\phi q}^{(3)}]_{21}^2 \nonumber\\
&+4.2[\WC_{\phi q}^{(1)}]_{11} +83.9[\WC_{\phi q}^{(1)}]_{11}^2 +8.8[\WC_{\phi q}^{(3)}]_{11}+83.9[\WC_{\phi q}^{(3)}]_{11}^2 \nonumber\\
&+0.3[\WC_{\phi q}^{(1)}]_{22} +3.7[\WC_{\phi q}^{(1)}]_{22}^2 +0.6[\WC_{\phi q}^{(3)}]_{22}+3.7[\WC_{\phi q}^{(3)}]_{22}^2 \nonumber\\
&+ 0.13[\WC_{\phi d}]_{21}+ 20.23 [\WC_{\phi d}]_{21}^2 \nonumber\\
&-18.1[\WC_{\phi q}^{(1)}]_{11}[\WC_{\phi q}^{(1)}]_{12} +1.3[\WC_{\phi q}^{(1)}]_{11}[\WC_{\phi q}^{(1)}]_{22} -8.1[\WC_{\phi q}^{(1)}]_{12}[\WC_{\phi q}^{(1)}]_{22} \nonumber\\
&-18.1[\WC_{\phi q}^{(3)}]_{11}[\WC_{\phi q}^{(3)}]_{12} +1.2[\WC_{\phi q}^{(3)}]_{11}[\WC_{\phi q}^{(3)}]_{22} -8.0[\WC_{\phi q}^{(3)}]_{12}[\WC_{\phi q}^{(3)}]_{22} \nonumber\\
&-50.1[\WC_{\phi q}^{(1)}]_{11}[\WC_{\phi q}^{(3)}]_{11} -18.3[\WC_{\phi q}^{(1)}]_{11}[\WC_{\phi q}^{(3)}]_{12} +1.2[\WC_{\phi q}^{(1)}]_{11}[\WC_{\phi q}^{(3)}]_{22} \nonumber\\
&-18.2[\WC_{\phi q}^{(1)}]_{12}[\WC_{\phi q}^{(3)}]_{11} -18.7[\WC_{\phi q}^{(1)}]_{12}[\WC_{\phi q}^{(3)}]_{12} -8.1[\WC_{\phi q}^{(1)}]_{12}[\WC_{\phi q}^{(3)}]_{22} \nonumber\\
&+1.3[\WC_{\phi q}^{(1)}]_{22}[\WC_{\phi q}^{(3)}]_{11}  -8.0[\WC_{\phi q}^{(1)}]_{22}[\WC_{\phi q}^{(3)}]_{12}  +3.1[\WC_{\phi q}^{(1)}]_{22}[\WC_{\phi q}^{(3)}]_{22}	\nonumber\\
&+0.06[\WC_{\phi q}^{(1)}]_{21}[\WC_{\phi d}]_{21}-0.044[\WC_{\phi q}^{(3)}]_{21}[\WC_{\phi d}]_{21} \nonumber\\
&-0.13[\WC_{\phi q}^{(1)}]_{11}[\WC_{\phi d}]_{21}-0.05[\WC_{\phi q}^{(3)}]_{11}[\WC_{\phi d}]_{21} \nonumber\\
&-0.022[\WC_{\phi q}^{(1)}]_{22}[\WC_{\phi d}]_{21}-0.016[\WC_{\phi q}^{(3)}]_{22}[\WC_{\phi d}]_{21} \nonumber\\
&+14.63[\WC_{dW}]_{21}^2+4.35[\WC_{dB}]_{21}^2+15.99[\WC_{dW}]_{21}[\WC_{dB}]_{21},
\label{WHpoly}
\end{align}

\section{Experimental results and fitted number of events for Monojet process} \label{appmj}

We use the inclusive measurements of the number of monojet events for different phase-space regions from the ATLAS analysis~\cite{ATLAS:2021kxv}. The predicted and observed number of events in the experiment are copied here in Table~\ref{tab:monojet_exp}. The fits to the number of monojet events as a function of WCs can be written in the form of Eq.~\ref{eq:N}. We present the values of the coefficients $\beta_i$ and $\gamma_{jk}$  for different phase-space regions in Table~\ref{tab:monojet_coeffs}. The errors corresponding to the estimated values of the coefficients are much smaller than the experimental errors and, therefore, ignored for our numerical estimates. 
\begin{table}[]
	\centering
			\begin{tabular}{c|c|c}\hline
				Region&Predicted&Observed\\ \hline 	\hline
				$\rm \MET>200$ GeV & 3\,120\,000 $\pm$ 40\,000 & 3\,148\,643 \\
				$\rm \MET>250$ GeV & 1\,346\,000 $\pm$ 16\,000 & 1\,357\,019 \\
				$\rm \MET>300$ GeV & 597\,000 $\pm$ 8000 & 604\,691 \\
				$\rm \MET>350$ GeV & 286\,000 $\pm$ 4000 & 290\,779 \\
				$\rm \MET>400$ GeV & 146\,400 $\pm$ 2300 & 149\,743 \\
				$\rm \MET>500$ GeV & 45\,550 $\pm$ 1000 & 46\,855 \\
				$\rm \MET>600$ GeV & 16\,800 $\pm$ 500 & 17\,397 \\
				$\rm \MET>700$ GeV & 7070 $\pm$ 240 & 7194 \\
				$\rm \MET>800$ GeV & 3180 $\pm$ 130 & 3208 \\
				$\rm \MET>900$ GeV & 1560 $\pm$ 80 & 1545 \\
				$\rm \MET>1000$ GeV & 720 $\pm$ 60 & 807 \\
				$\rm \MET>1100$ GeV & 407 $\pm$ 34 & 394 \\
				$\rm \MET>1200$ GeV & 223 $\pm$ 19 & 207 \\
				\hline	\hline
			\end{tabular}
\caption{ \small Data and SM background predictions for different inclusive $\MET$-bins in the experimental monojet analysis~\cite{ATLAS:2021kxv}.
}
	\label{tab:monojet_exp}
\end{table}

\begin{table}[]
	\centering
		\begin{tabular}{ c|c|c|c|c|c|c|c|c|c|c|c|c|c}
			\hline
			Combinations & 200 & 250 & 300 & 350 & 400 & 500 & 600 & 700 & 800 & 900 & 1000 & 1100 & 1200\\ \hline
			\hline
			$[\wclqone]_{2212}$		& 1365 & 859 & 491 & 286 & 167 & 81 & 38 & 18 & 8.7 & 4.7 & 2.7 & 1.6  & 0.9\\ \hline
			$[\wclqthree]_{2212}$	& -1366 & -861 & -497 & -286 & -174 & -82 & -38 & -18 & -9.0 & -4.7 & -2.7 & -1.6 & -0.9\\ \hline
			$[\wcphiqone]_{12}$	& -29458 & -13770 & -6201 & -2971 & -1453 & -520 & -202 & -83 & -3.7 & -1.5 & -8.3 & -4.5 & -2.5\\ \hline
			$[\wcphiqthree]_{12}$	& -28410 & -12960 & -6053 & -3096 & -1507 & -505 & -199 & -80 & -3.4 & -1.5 & -8.3 & -4.5 & -2.5\\ \hline
			$([\wclqone]_{2212})^2$	& 14269 & 10856 & 7597 & 5201 & 3590 & 1665 & 946 & 536 & 312 & 168 & 109 & 68 & 43 \\ \hline
			$([\wclqthree]_{2212})^2$	& 14177 & 10639 & 7389 & 5133 & 3566 & 1681 & 963 & 547 & 309 & 167 & 108 & 68 & 43 \\ \hline
			$([\wcphiqone]_{12})^2$	& 8951 & 4185 & 1899 & 905 & 436 & 127 & 49 & 20 & 9.2 & 3.5 & 1.9 & 1 & 0.6\\ \hline
			$([\wcphiqthree]_{12})^2$	& 8954 & 4168 & 1924 & 919 & 471 & 128 & 50 & 21 & 9.2 & 3.4 & 1.9 & 1 & 0.6\\ \hline
			$([\wcphid]_{12})^2$	& 140 & 70 & 35 & 17 & 9 & 2.5 & 1 & 0.42 & 0.2 & 0.1 & 0.04 & 0.02 & 0.01\\ \hline
			$([\wcld]_{2212})^2$	& 6094  & 4460 & 3010 & 2077 & 1377 & 600 & 334 & 180 & 102 & 60 & 34 & 21 & 13\\ \hline
			$([\WC_{dZ}]_{12})^2$	& 126361 & 76658 & 45966 & 30004 & 18127 & 8598 & 4312 & 2314 & 1256 & 648 & 401 & 245 & 151\\ \hline\hline
            $[\wclqone]_{2211}$		& 1695 & 1090 & 659 & 414 & 251 & 103 & 44 & 35 & 22 & 13 & 7 & 4  & 3\\ \hline
		$[\wclqthree]_{2211}$	& 9246 & 6026 & 3660 & 2212 & 1370 & 586 & 263 & 138 & 80 & 45 & 26 & 16 & 10\\ \hline
		$[\wcphiqone]_{11}$ & -36546 & -20650 & -10974 & -10007 & -9228 & -2096 & -1480 & -113 & -76 & -43 & -30 & -16 & -8\\ \hline
		$[\wcphiqthree]_{11}$	& 250267 & 122751 & 59066 & 30505 & 18052 & 6437 & 2366 & 632 & 330 & 172 & 90 & 48 & 22\\ \hline
		$([\wclqone]_{2211})^2$	& 19209 & 14541 & 10106 & 7136 & 4765 & 2690 & 1611 & 768 & 491 & 303 & 189 & 120 & 76 \\ \hline
		$([\wclqthree]_{2211})^2$ & 19380 & 14241 & 10185 & 7153 & 5159 & 2519 & 1386 & 775 & 498 & 307 & 189 & 122 & 81 \\ \hline
		$([\wcphiqone]_{11})^2$	& 9212 & 4375 & 1989 & 983 & 542 & 203 & 72 & 23 & 11 & 6 & 3 & 2 & 1\\ \hline
		$([\wcphiqthree]_{11})^2$	& 9177 & 4460 & 2060 & 1002 & 527 & 158 & 65 & 23 & 12 & 6 & 3 & 2 & 1\\ \hline
			\hline
            $[\wclqone]_{2222}$		& -357 & -217 & -122 & -69 & -39 & -15 & -6 & -4 & -2 & -1 & -0.7 & -0.4  & -0.3\\ \hline
		$[\wclqthree]_{2222}$	& 1114 & 668 & 372 & 199 & 109 & 45 & 19 & 10 & 6 & 3 & 2 & 1 & 0.6\\ \hline
		$[\wcphiqone]_{22}$		& 12445 & 5997 & 3334 & 1401 & 853 & 568 & 67 & 19 & 10 & 4 & 2 & 1.3 & 1\\ \hline
		$[\wcphiqthree]_{22}$	& 44514 & 21077 & 10122 & 3968 & 1550 & 338 & 130 & 56 & 25 & 14 & 7 & 3 & 2\\ \hline
		$([\wclqone]_{2222})^2$	  & 1477 & 1055 & 728 & 493 & 330 & 298 & 118 & 40 & 24 & 14 & 8 & 5 & 3 \\ \hline
		$([\wclqthree]_{2222})^2$ & 1482 & 1056 & 727 & 498 & 334 & 297 & 115 & 40 & 24 & 14 & 8 & 5 & 3\\ \hline
		$([\wcphiqone]_{22})^2$	    & 1446 & 635 & 264 & 115 & 57 & 16 & 4 & 2 & 1 & 0.4 & 0.2 & 0.1 & 0.06\\ \hline
		$([\wcphiqthree]_{22})^2$	& 1490 & 689 & 312 & 142 & 60 & 12 & 5 & 2 & 1 & 0.4 & 0.2 & 0.1 & 0.06\\ \hline
        \hline
	\end{tabular}
	\caption{\small Pure linear and quadratic coefficients $\beta_i$ and $\gamma_{jj}$ (Eq.~\ref{eq:N}) for different combinations of WCs that contributes to the monojet process. The columns indicates different inclusive bins used in the experimental analysis~\cite{ATLAS:2021kxv}.  Entries below $0.05\%$ of the largest entry are dropped.}
	\label{tab:monojet_coeffs}
\end{table}

\begin{table}[]
	\centering
	\begin{tabular}{ c|c|c|c|c|c|c|c|c|c|c|c|c|c}
		\hline
		Combinations & 200 & 250 & 300 & 350 & 400 & 500 & 600 & 700 & 800 & 900 & 1000 & 1100 & 1200\\ \hline
		\hline
		$[\wclqone]_{2212}$ - $[\wclqthree]_{2212}$	& 4747 & 4005 & 2911 & 2231 & 1632 & 765 & 458 & 275 & 168 & 106 & 72 & 46 & 30\\ \hline
			$[\wclqone]_{2212}$ - $[\wcphiqone]_{12}$	& -701 & -463 & -260 & -160 & -105 & -40 & -20 & -8.7 & -4 & -2.0 & -1.5 & -0.9 & -0.4\\ \hline
			$[\wcphiqone]_{12}$ - $[\wcphiqthree]_{12}$	& -1661 & -883 & -497 & -247 & -141 & -44 & -18 & 8 & -4.5 & -1.7 & -1 & -0.57 & -0.3\\ \hline
			$[\wcld]_{2212}$ - $[\wcphid]_{12}$	& -261 & -150 & -64 & -32 & -14 & -8 & -4 & -1.3 & -0.88 & -0.3 & -0.2 & -0.1 & 0.08\\ \hline
		$[\wclqone]_{2211}$ - $[\wclqthree]_{2211}$	& 16100 & 12340 & 9258 & 6862 & 4616 & 2782 & 1513 & 756 & 516 & 327 & 210 & 139 & 95\\ \hline
		$[\wclqone]_{2211}$ - $[\wcphiqone]_{11}$	& -1854 & -967 & -619 & -346 & -140 & -48 & -69 & -11 & -4 & -3 & -1.5 & -0.06 & -0.02\\ \hline
        $[\wclqone]_{2222}$ - $[\wclqthree]_{2222}$	& -953 & -689 & -496 & -345 & -234 & -180 & -120 & -33 & -21 & -12 & -8 & -5 & -3 \\ \hline
		$[\wclqone]_{2222}$ - $[\wcphiqone]_{22}$	& -68 & -46 & -28 & -18 & -10 & -5 & -2 & -1 & -0.5 & -0.4 & -0.3 & -0.2 & -0.1\\ \hline
		\hline
		$[\wcphiqone]_{11}$ - $[\wcphiqthree]_{11}$	& -4766 & -2575 & -1213 & -613 & -346 & -136 & -51 & -16 & -9 & -5 & -3 & -2 & -1\\ \hline
		$[\wcphiqone]_{11}$ - $[\wcphiqthree]_{12}$& -1467 & -742 & -342 & -114 & -54 & -25 & -10 & -4 & -2 & -1 & -0.6 & -0.3 & -0.13\\ \hline
		$[\wcphiqone]_{12}$ - $[\wcphiqthree]_{11}$ & -1284 & -578 & -292 & -129 & -111 & -66 & -7 & -3.4 & -2 & -1 & -0.5 & -0.34 & -0.2\\ \hline
		$[\wcphiqone]_{12}$ - $[\wcphiqthree]_{22}$	 & -1089 & -514 & -210 & -83 & -31 & -12 & -4 & -3 & -1.3 & -0.6 & -0.3 & -0.2 & -0.1 \\ \hline
		$[\wcphiqone]_{12}$ - $[\wcphiqthree]_{12}$	& -1661 & -883 & -497 & -247 & -141 & -44 & -18 & 8 & -4.5 & -1.7 & -1 & -0.57 & -0.3\\ \hline
		$[\wcphiqone]_{22}$ - $[\wcphiqthree]_{12}$	 & -1003 & -461 & -258 & -74 & -36 & -14 & -5 & -3 & -1.4 & -0.8 & -0.4 & -0.2 & -0.1 \\ \hline
		$[\wcphiqone]_{22}$ - $[\wcphiqthree]_{22}$	& -996 & -456 & -208 & -116 & -97 & -70 & -50 & -33 & -21 & -12 & -8 & -5 & -3 \\ \hline
		\hline
		$[\wclqone]_{2211}$ - $[\wclqthree]_{2211}$	& 13202 & 10118 & 7591 & 5626 & 3785 & 2281 & 1240 & 619 & 422 & 268 & 172 & 113 & 77\\ \hline
		$[\wclqone]_{2211}$ - $[\wclqthree]_{2212}$	& 3412 & 2408 & 1736 & 1134 & 795 & 451 & 246 & 123 & 65 & 40 & 23 & 14 & 9\\ \hline
		$[\wclqone]_{2211}$ - $[\wclqthree]_{2222}$	& -401 & -330 & -235 & -150 & -89 & -45 & -20 & -7 & -4 & -3 & -1.9 & -1.3 & -0.6\\ \hline
		$[\wclqone]_{2212}$ - $[\wclqthree]_{2211}$	& 3609 & 2489 & 1910 & 1440 & 946 & 399 & 205 & 113 & 77 & 49 & 35 & 18 & 10\\ \hline
		$[\wclqone]_{2212}$ - $[\wclqthree]_{2222}$	& 2324 & 1792 & 1279 & 774 & 605 & 312 & 164 & 80 & 51 & 31 & 21 & 12 & 6\\ \hline
		$[\wclqone]_{2212}$ - $[\wclqthree]_{2212}$	& 4747 & 4005 & 2911 & 2231 & 1632 & 765 & 458 & 275 & 168 & 106 & 72 & 46 & 30\\ \hline
		$[\wclqone]_{2222}$ - $[\wclqthree]_{2211}$	& -409 & -331 & -240 & -152 & -94 & -44 & -20 & -8 & -6 & -5 & -3 & -1.9 & -0.6\\ \hline
		$[\wclqone]_{2222}$ - $[\wclqthree]_{2212}$ & 2183 & 1590 & 1166 & 857 & 543 & 295 & 154 & 86 & 51 & 29 & 19 & 12 & 6 \\ \hline
		$[\wclqone]_{2222}$ - $[\wclqthree]_{2222}$	& -953 & -689 & -496 & -345 & -234 & -180 & -120 & -33 & -21 & -12 & -8 & -5 & -3 \\ \hline
		\hline
	\end{tabular}
 	\caption{\small  Same as Fig~\ref{tab:monojet_coeffs} but for the coefficients $\gamma_{jk}$ (Eq.~\ref{eq:N}) of NP-NP interference terms for different combinations of WCs.}
	\label{tab:monojet_coeffs2}
\end{table}

\section{Widths Z and W boson}
\label{app:Width_calculations}
The operators affecting Z and W boson widths are:
\begin{align}
	\Op_{d W} &= \left( \bar q \sigma^{\mu \nu} d \right) \tau^I \varphi W_{\mu \nu}^I\nonumber\\
	\Op_{\varphi q}^{(1)}&=\left( \varphi^\dagger i \overleftrightarrow{D}_\mu \varphi \right) \left( \bar q \gamma^\mu q \right)\nonumber\\
	\Op_{\varphi q}^{(3)} &=\left( \varphi^\dagger i\tau^I \overleftrightarrow{D}_\mu \varphi \right) \left( \bar q \tau^I \gamma^\mu q \right) ,\nonumber\\
	\Op_{\varphi d}  &= \left( \varphi^\dagger i D_\mu \varphi \right) \left( \bar d \gamma^\mu d \right) ,\nonumber\\
\end{align}

We consider tree-level correction up to quadratic order in EFT.
\subsection{Z-width}
The interaction of the Z-boson with a fermion f is:
\begin{align}
	\mathcal{L}=\frac{g}{cos \theta_W} Z_{\mu} \bar{f}\gamma^{\mu} \left(g_{L}^f P_L+g_R^f P_R\right) f
\end{align}	

Where, at tree level,
\begin{align}
	g_R^{f,SM} =-s_W^2 Q_f\quad{\rm and}\quad g_L^{f,SM}=T_3^f -s_W^2 Q_f.
\end{align}

We define the SMEFT contributions to the couplings as $\delta g_{L}^f$ and $\delta g_{R}^f$; where $\delta g^f = g^{f} - g^{f,SM}$. 

Therefore, the SMEFT contributions can be written as:
\begin{align}
	\delta g_L^{u}&= -\frac{v^2}{2\Lambda^2}\left(\WC_{\phi q}^{(1)}-\WC_{\phi q}^{(3)}\right)\\
	\delta g_L^{d} &= -\frac{v^2}{2\Lambda^2}\left(\WC_{\phi q}^{(1)}+\WC_{\phi q}^{(3)}\right) \\
	\delta g_R^{u}&= -\frac{v^2}{2\Lambda^2} \WC_{\phi u} \;(\to 0, \;\textrm{for this work})\\
	\delta g_R^{d} &= -\frac{v^2}{2\Lambda^2} \WC_{\phi d}
\end{align}

Using a notation $\frac{G_F M_Z^3}{6\sqrt{2} \pi}=FF\simeq0.33$, the total decay width of the Z-boson is, therefore,

\begin{align}
	\Gamma_Z^{tot} &= FF\sum_{f} N_c^f \left[\left(g_{L}^f+\delta g_{L}^f\right)^2 + \left(g_{R}^f+\delta g_{R}^f\right)^2\right]\nonumber\\
	&=  \Gamma_Z^{SM} + FF\sum_{f} N_c^f  \left[2 g_{L}^f\delta g_{L}^f + 2 g_{R}^f \delta g_{R}^f \right] +FF\sum_{f} N_c^f  \left[(\delta g_{L}^f)^2 + (\delta g_{R}^f)^2 \right]
\end{align}
\begin{align}
	\Gamma_{Z,had}^{tot} &=  \Gamma_Z^{SM} + \underbrace{FF\left[\sum_{u,c}  6\left(g_{L}^u\delta g_{L}^u \right) + \sum_{d,s,b}  6(g_{L}^d\delta g_{L}^d + g_{R}^d \delta g_{R}^d)\right]}_{\rm flavor-diagonal} + \underbrace{FF\sum_{f}  3\left[(\delta g_{L}^f)^2 + (\delta g_{R}^f)^2 \right]}_{\rm diagonal\,+ \, off-diagonal}\nonumber\\
	&=  \Gamma_Z^{SM} - \frac{FF\, v^2}{\Lambda^2}\left[\sum_{\substack{u_1,u_2\in 1,2\\ u_1=u_2}}6(\frac{1}{2}-\frac{2}{3}s_W^2) (\wcphiqone-\wcphiqthree)_{u1u2}\right] \nonumber\\
    &~~~~~~~~~~ -\frac{FF\, v^2}{\Lambda^2} \left[\sum_{\substack{d_1,d_2\in 1,2,3\\ d_1=d_2}}6(-\frac{1}{2}+\frac{1}{3}s_W^2)(\wcphiqone+\wcphiqthree)_{d_1d_2} + \frac{6}{3}s_W^2 (\wcphid)_{d_1d_2} \right] \nonumber\\
    &~~~~~~~~~~~ - \frac{FF v^2}{\Lambda^2}\left[-12s_\theta c_\theta(-\frac{1}{2}+\frac{1}{3}s_W^2)(\wcphiqone+\wcphiqthree)_{12} - \frac{12}{3}s_\theta c_\theta s_W^2 (\wcphid)_{12}\right]\nonumber\\
	& ~~~~~~ + \frac{FF v^4}{\Lambda^4} \sum_{\substack{u_1,u_2,d_1,d_2\\u_1,u_2\in 1,2\\d_1,d_2\in 1,2,3}}3\left[(\wcphiqone-\wcphiqthree)_{u_1u_2}^2 + (\wcphiqone+\wcphiqthree)_{d_1d_2}^2 +  (\wcphid)_{d_1d_2}^2 \right]
    \end{align}
\begin{align}
	&=  \Gamma_Z^{SM} + \frac{v^2}{\Lambda^2}\left[\sum_{\substack{u_1,u_2\in 1,2\\ u_1=u_2}} -0.69\,\left([\wcphiqone]_{u_1u_2} -[\wcphiqthree]_{u_1u_2}\right)\right] \nonumber\\
    &~~~~~~~~~~+\frac{v^2}{\Lambda^2}\left[\sum_{\substack{d_1,d_2\in 1,2,3\\ d_1=d_2}}(0.85([\wcphiqone]_{d_1d_2}+[\wcphiqthree]_{d_1d_2}) -0.14 [\wcphid]_{d_1d_2}) \right] \nonumber\\
    &~~~~~~~~~ + \frac{v^2}{\Lambda^2}\left[-0.37(\wcphiqone+\wcphiqthree)_{12} + 0.065\,(\wcphid)_{12}\right]\nonumber\\
	& ~~~~~~~~ + \frac{0.99 \,v^4}{\Lambda^4} \left[2[\wcphiqone]_{12}^2+2[\wcphiqthree]_{12}^2 +[\wcphid]_{12}^2\right]  + \frac{0.99 \,v^4}{\Lambda^4} \sum_{kl\in 11,22,13,23}\left[2[\wcphiqone]_{kl}^2+2[\wcphiqthree]_{kl}^2 +[\wcphid]_{kl}^2\right]\nonumber\\
\end{align}

Note that, considering up-diagonal basis in third row of Eq. C9, only dd and ss combinations appear from the mixing of the down-quarks that interfere with SM. Since we are interested in only the $[\WC]_{12,11,22}$ operators affecting Kaon observables, we set other possibilities to zero, and use:
\begin{align}
	\Gamma_Z^{tot}&= \Gamma_Z^{SM} + (-0.022([\wcphiqone]_{12}+[\wcphiqthree]_{12})+0.0036[\wcphid]_{12})\nonumber\\
    &+ 0.01 \left([\wcphiqone]_{11}+[\wcphiqone]_{22}\right) + 0.09 \left([\wcphiqthree]_{11}+[\wcphiqthree]_{22}\right) -0.008 \left([\wcphid]_{11} + [\wcphid]_{22}\right)\nonumber\\
    &+10^{-2}\left(0.72\,([\wcphiqone]_{12}^2+[\wcphiqone]_{11}^2+[\wcphiqone]_{22}^2) + 0.72\, ([\wcphiqthree]_{12}^2+[\wcphiqthree]_{11}^2+[\wcphiqthree]_{22}^2)\right.\nonumber\\
    &\left.+  0.36 \, ([\wcphid]_{12}^2 +[\wcphid]_{11}^2 + [\wcphid]_{22}^2) \right)
\end{align}
\subsection{W-width}
The interaction of the Z-boson with a fermion f is:
\begin{align}
	\mathcal{L}=\frac{g_L^W}{\sqrt{2}} W_{\mu}^+ \bar{u}\gamma^{\mu} P_L d
\end{align}	

Therefore, the SMEFT contribution to the hadronic part of  $g_L^W$ is:
\begin{align}
	\delta g_{L}^{Wq} = \frac{v^2}{\Lambda^2} \wcphiqthree 
\end{align}

The total decay width of the W-boson is, therefore,
\begin{align}
	\Gamma_W &= \frac{G_F M_W^3}{3\sqrt{2} \pi}\sum_{f} N_c^f \left(1+\frac{\wcphiqthree v^2}{\Lambda^2}\right)^2\nonumber\\
	&= \Gamma_W^{SM} + \frac{G_F M_W^3}{3\sqrt{2} \pi}\sum_{f} N_c^f \left[2\frac{\wcphiqthree v^2}{\Lambda^2} + \left(\frac{\wcphiqthree v^2}{\Lambda^2}\right)^2\right]\nonumber\\
	\Gamma_W^{had}	&= \Gamma_W^{SM, had} + \frac{G_F M_W^3 v^2}{\sqrt{2} \pi \Lambda^2}\sum_{\substack{i=u,c;\\j=d,s,b}}|V_{ij}|^2\left[2[\wcphiqthree]_{ij} + \left(\frac{v^2}{\Lambda^2}\right)[\wcphiqthree]_{ij}^2\right]\nonumber\\
\Gamma_W^{had}	&= \Gamma_W^{SM, had} + \frac{v^2}{\Lambda^2} \left[2.72\,\left(|V_{11}|^2 [\wcphiqthree]_{11}+|V_{22}|^2 [\wcphiqthree]_{22}+|V_{12}|^2 [\wcphiqthree]_{12}\right)\right.\nonumber\\
&\left. \quad\quad\quad\quad\quad\quad ~~~ +0.16 \, \left(|V_{11}|^2 [\wcphiqthree]_{11}^2+|V_{22}|^2 [\wcphiqthree]_{22}^2+|V_{12}|^2[\wcphiqthree]_{12}^2\right)\right]\nonumber\\ 
& ~~~~~+\frac{v^2}{\Lambda^2}\sum_{\substack{kl\in ub,cb}}|V_{kl}|^2\left[2.72\,[\wcphiqthree]_{kl} + 0.16[\wcphiqthree]_{kl}^2\right]\nonumber\\
\end{align}

Therefore, the contribution relevant to us is:
\begin{align}
	\Gamma_W^{tot}	&= \Gamma_W^{SM} +0.008 \, [\wcphiqthree]_{12}+ 0.0004\, [\wcphiqthree]_{12}^2++0.16 \, \left([\wcphiqthree]_{11}+[\wcphiqthree]_{22}\right)+ 0.009\, \left([\wcphiqthree]_{11}^2+[\wcphiqthree]_{22}^2\right).
\end{align}

\section{Individual bounds on SMEFT operators}
We present the bounds on individual SMEFT operators in Table~\ref{tab:individual_SMEFT}. These bounds essentially represent one-dimensional slices of the combined $95\%$ CL allowed ranges of the 24 SMEFT WCs, where all but one WC are set to their minimum values, one at a time.
\begin{table}[]
	\centering
    \caption{\small Slices of the combined $95\%$ CL allowed ranges for individual SMEFT WCs taken through the minimum for the current and projected luminosities.}
	\begin{tabular}{ |c|c|c|c|c||c|c|c|c|c|}
		\hline
\multicolumn{5}{|c||}{Muon} & \multicolumn{5}{c|}{Electron}\\ \hline
        Luminosity $\to$& \multicolumn{2}{c}{$140\;fb^{-1}$}  &\multicolumn{2}{|c||}{$3000\;fb^{-1}$}& Luminosity $\to$ &\multicolumn{2}{c}{$140\;fb^{-1}$}  &\multicolumn{2}{|c|}{$3000\;fb^{-1}$}  \\\hline
\hline
Parameter & Min & Max 		& Min & Max & Parameters & Min & Max 		& Min & Max 		\\ \hline
$\wclqoneMOO$ & -0.017 & 0.042 	& -0.005 & 0.017 & $\wclqoneEOO$ & -0.022 & 0.040 & -0.004 & 0.019\\ \hline
$\wclqoneMTT$ & -0.169 & 0.170 	& -0.100 & 0.054 & $\wclqoneETT$ & -0.283 & 0.163 & -0.101 & 0.051\\ \hline
$\wclqoneMOT$ & -0.049 & 0.032 	& -0.024 & 0.016 & $\wclqoneEOT$ & -0.040 & 0.047 & -0.021 & 0.017\\ \hline
$\wclqthreeMOO$ & -0.003 & 0.006 & -0.001 & 0.001 & $\wclqthreeEOO$ & -0.009 & 0.000 & -0.001 & 0.001\\ \hline
$\wclqthreeMTT$ & -0.136 & 0.047 & -0.029 & 0.011 & $\wclqthreeETT$ & -0.013 & 0.089 & -0.025 & 0.008	\\ \hline
$\wclqoneMOT$ & -0.030 & 0.017 	& -0.013 & 0.005 & $\wclqthreeEOT$ & -0.025 & 0.013 & -0.010 & 0.005	\\ \hline
$\wcqeMOO$ & -0.025 & 0.037 	& -0.010 & 0.017 & $\wcqeEOO$ & -0.028 & 0.037 & -0.009 & 0.017	\\ \hline
$\wcqeMTT$ & -0.160 & 0.178 	& -0.069 & 0.087 & $\wcqeETT$ & -0.171 & 0.203 & -0.068 & 0.087 	\\ \hline
$\wcqeMOT$ & -0.041 & 0.041 	& -0.017 & 0.020 & $\wcqeEOT$ & -0.047 & 0.041 & -0.018 & 0.019	\\ \hline
$\wcphiqoneOO$ & -0.223 & 0.152 	& -0.108 & -0.006& $\wcphiqoneOO$ & -0.199 & 0.120 & -0.103 & -0.021	\\ \hline
$\wcphiqoneTT$ & -0.924 & 0.870 	& -0.240 & 0.298 & $\wcphiqoneTT$ & -0.785 & 0.752 & -0.173 & 0.260	\\ \hline
$\wcphiqoneOT$ & -0.217 & 0.255 	& -0.058 & 0.088 & $\wcphiqoneOT$ & -0.184 & 0.221 & -0.047 & 0.072	\\ \hline
$\wcphiqthreeOO$ & -0.163 & 0.121 	& -0.010 & 0.047 & $\wcphiqthreeOO$ & -0.123 & 0.106 & -0.006 & 0.036 	\\ \hline
$\wcphiqthreeTT$ & -0.835 & 0.753 	& -0.175 & 0.290 & $\wcphiqthreeTT$ & -0.719 & 0.666 & -0.116 & 0.236	\\ \hline
$\wcphiqthreeOT$ & -0.185 & 0.222 	& -0.057 & 0.087 & $\wcphiqthreeOT$ & -0.166 & 0.202 & -0.041 & 0.069 	\\ \hline
$\wcldMOT$ & -0.068 & 0.068 		& -0.031 & 0.031 & $\wcldEOT$ & -0.074 & 0.074 & -0.031 & 0.031	\\ \hline
$\wcedMOT$ & -0.068 & 0.068 		& -0.031 & 0.031 & $\wcedEOT$ & -0.073 & 0.073 & -0.031 & 0.031 	\\ \hline
$\wcledqMOO$ & -0.036 & 0.036 	& -0.016 & 0.016 & $\wcledqEOO$ & -0.032 & 0.032 & -0.015 & 0.015	\\ \hline
$\wcledqMOT$ & -0.071 & 0.071 	& -0.032 & 0.032 & $\wcledqEOT$ & -0.071 & 0.071 & -0.031 & 0.031 	\\ \hline
$\wcledqMTO$ & -0.058 & 0.058 	& -0.026 & 0.026 & $\wcledqETO$ & -0.050 & 0.050 & -0.023 & 0.023 	\\ \hline
$\wcphidOT$ & -0.391 & 0.385 		& -0.123 & 0.117 & $\wcphidOT$ & -0.337 & 0.330 & -0.101 & 0.095	\\ \hline
$[\wcdgamma]_{12}$ & -0.906 & 0.906	& -0.440 & 0.440 & $[\wcdgamma]_{12}$ & -0.793 & 0.793  & -0.401 & 0.401	\\ \hline
$[\wcdgamma]_{21}$ & -0.842 & 0.880 &  -0.466 & 0.462  & $[\wcdgamma]_{21}$ & -0.815 & 0.819 & -0.426 & 0.433	\\ \hline
$[\wcdgamma]_{11}$ & -0.578 & 0.528 & -0.283 & 0.234 & $[\wcdgamma]_{11}$ & -0.529 & 0.511 & -0.263 & 0.215	\\ \hline
\end{tabular}
\label{tab:individual_SMEFT}
\end{table}

\bibliographystyle{JHEP}
\bibliography{dilepton_LEFT.bib}
\end{document}